\documentclass{aa}

\usepackage{graphicx}
\usepackage{amsmath}	
\usepackage{amssymb}	
\usepackage{relsize}
\usepackage{txfonts}
\usepackage[colorlinks=true,linkcolor=blue,allcolors=blue]{hyperref}%
\usepackage[switch]{lineno}

\newcommand{\Msun}{M$_\odot$}

\begin{document}

   \title{The diverse physical origins of stars in the dynamically hot bulge: CALIFA vs. IllustrisTNG}


   \author{Le Zhang\inst{1,2}\thanks{E-mail: zhangle@shao.ac.cn},
          Ling Zhu\inst{1}\thanks{Corr author: lzhu@shao.ac.cn},
          Annalisa Pillepich\inst{3},
          Min Du\inst{4},
          Fangzhou Jiang\inst{5},
          \and
          Jes{\'u}s Falc{\'o}n-Barroso\inst{6,7}}

   \institute{$^{1}$Shanghai Astronomical Observatory, Chinese Academy of Sciences, 80 Nandan Road, Shanghai 200030, China\\
             $^{2}$School of Astronomy and Space Sciences, University of Chinese Academy of Sciences, No. 19A Yuquan Road, Beijing 100049, People’s Republic of China\\
             $^{3}$Max-Planck-Institut f{\"u}r Astronomie, K{\"o}nigstuhl 17, D-69117 Heidelberg, Germany\\
             $^{4}$Department of Astronomy, Xiamen University, Xiamen, Fujian 361005, China\\
             $^{5}$Kavli Institute for Astronomy and Astrophysics, Peking University, Beijing 100871, China\\
             $^{6}$Instituto de Astrof\'isica de Canarias, Calle Via L\'{a}ctea s/n, 38200 La Laguna, Tenerife, Spain \\
             $^{7}$Depto. Astrof\'isica, Universidad de La Laguna, Calle Astrof\'isico Francisco S\'{a}nchez s/n, 38206 La Laguna, Tenerife, Spain \\}
             
\titlerunning{Physical origin of hot bulge}
\authorrunning{Le Zhang et al.}
 
  \abstract
   {We compare the internal stellar structures of central galaxies in the TNG50 and TNG100 simulations and field galaxies in the CALIFA survey. The luminosity fractions of the dynamically cold, warm, and hot components in both TNG50 and TNG100 galaxies exhibit general consistency with those observed in CALIFA galaxies. For example, they all exhibit a minimum luminosity fraction ($f_{\rm hot} \sim$ 0.18) of the dynamically hot component in galaxies with stellar masses of $M_*\sim 1-2 \times 10^{10}$\ \Msun, and the morphology of each orbital component in the TNG50 and TNG100 galaxies closely resembles that found in the CALIFA galaxies. We therefore use the simulations to quantify the physical origins of the different components, focusing on the dynamically hot component in TNG50. We identify three primary regimes and thus physical processes: (1) in low mass galaxies ($M_*\lesssim 10^{10}$\ \Msun) that have not experienced major mergers, stars are born with a wide range of circularity distributions and have remained relatively unchanged until the present day. Consequently, hot stars in such galaxies at redshift $z = 0$ are predominantly born hot. (2) In higher mass galaxies ($M_*\gtrsim 10^{10}$\ \Msun) lacking major mergers, most stars are initially born cold but are subsequently heated through secular evolution. (3) In galaxies across the entire mass range, mergers, if they occurred, significantly increased the hot orbital fraction. As a result, the dynamically hot bulge within $R_e$ of present-day galaxies does not indicate their past merger histories; instead, the hot stars in the outer regions are mostly heated or accreted by mergers, thus indicating galaxy merger history. Massive galaxies are initially born with cold, rotationally supported structures, consistent with recent observations from the James Webb Space Telescope (JWST) regarding high-redshift galaxies.
    }

   \maketitle
   

\section{Introduction}

Galaxies consist of various structural components, including thin disks, thick disks, classical bulges, pseudo-bulges, and stellar halos, each of which serves as a historical record of their formation and assembly. Stars in thin disks form and remain in quiescent environments. However, the formation of bulges involves several proposed mechanisms:
(1) stars may be heated from rotation-supported structures through mergers with other galaxies \citep[e.g.][]{Davies1983, Quinn1993} or via secular evolution \citep[e.g.][]{Jenkins1990, Aumer2016};
(2) bulges could also arise from the compact and rapid formation of central gaseous clumps carried by filaments and gas-rich mergers or formed from violent disk instability, especially during high redshift periods \citep[e.g.][]{Elmegreen2008, lapiner2023wet};
(3) stars in non-circular orbits might originate, in part, from the accretion of satellite galaxies \citep[e.g.,][]{Facundo2017, Read2008}.
Galaxies do not exist in isolation; their formation and evolution are intertwined with larger-scale environments. Therefore, interactions with other galaxies and mergers can profoundly affect or even determine the distribution of stellar orbits in galaxies. This is particularly evident in the case of stars that make up stellar halos in the Milky Way (MW). \citep{Helmi2018}. 

Modern hydrodynamical cosmological simulations, such as IllustrisTNG \footnote{https://www.tng-project.org} and EAGLE \footnote{http://eagle.strw.leidenuniv.nl/}, have demonstrated the ability to generate a variety of galaxy types. These simulations offer a unique opportunity to quantitatively investigate the origins of different galaxy structures within a cosmic volume.
In cosmological simulations, galaxies exhibit diverse merger histories. Major mergers, if experienced by a galaxy, have the potential to disrupt disks and lead to substantial growth in bulges and halos \citep{Sotillo-Ramos2022, Zhu2022b, Pulsoni2021, Lagos2018}. However, intriguingly, there are galaxies dominated by bulges that appear to have been minimally affected by mergers since $z\sim 2$ \citep{du2021evolutionary}, suggesting alternative pathways for bulge growth. 
The secular evolution \citep{Seidel2015} and the dissolution of the bar by the infalling gas might play an important role in the formation of the bulge \citep{bournaud2002gas, bournaud2005lifetime, du2017orthogonal, guo2020new,li2023nested}. Disk heating by secular evolution plays a significant role in creating the dynamically hot bulge in half of the quiescent elliptical galaxies in TNG50 \citep{park2022formation} and in MW-like galaxies from the Auriga simulation \citep{Fragkoudi2020}.
However, in some other zoom-in MW-like simulations \citep{Yu2022, Bird2013, Stinson2013}, their structures, including thin disks, thick disks, and dynamically hot bulges, evolve in a manner that aligns with their formation at birth. Bulges are formed by stars born dynamically hot; neither merger nor secular evolution plays a significant role in their formation.
The formation of bulges might depend on the galaxy formation model, e.g. feedback process \citep{Obreja2013}, as well as the galaxy stellar mass, on which merger frequency and bar frequency strongly depend. The relative contributions of these diverse physical processes to the formation of bulges in a representative sample of galaxies, spanning the Hubble sequence and a broad range of masses, have yet to be quantified.

On the one hand, simulations require calibration against actual observed galaxies. However, the task of comparing galaxy structures derived from observations with those from simulations is not straightforward. In the simulations mentioned earlier, the definitions of structures such as disks, bulges, and halos are typically dynamic and can vary between different studies. Conversely, in the case of real galaxies, morphological decomposition techniques, such as the bulge-disk decomposition, have been extensively employed, yielding invaluable insights over the past decades. However, it is important to note that these methods are subject to significant uncertainties, which makes them challenging to compare fairly with the simulation results. It is only through the establishment of systematic and well-defined methodologies for comparing observed and simulated galaxies that substantial progress in this field can be achieved.

The Milky Way stands as the only real galaxy amenable to dynamic decomposition in a manner comparable to simulations, mainly due to the availability of 6D phase-space information for its stars, as gleaned from observations (e.g. recently \citealt{Helmi2018, Belokurov2018}).
Over the past decade, integral field unit (IFU) spectroscopy surveys, including SAURON \citep{davies2001}, SAMI \citep{croom2012}, CALIFA \citep{sanchez2012}, MaNGA \citep{bundy2014}, have provided stellar kinematic maps for thousands of galaxies. With an orbit superposition model \cite{schwarzschild1979} for IFU data, insights of the internal distribution of stellar orbits become possible for nearby galaxies, with some approximations.
The method \citep{vdB2008} has been widely applied to large galaxies samples observed by SAURON \citep{cappellari2016}, CALIFA \citep{zhu2018nature}, MaNGA \citep{Jin2020}, SAMI \citep{Santucci2022} and MUSE \citep{Ding2023, Zhu2022b, Poci2021}. Recently, the code has been made publicly available \citep{Thater2022} and has been enhanced to explicitly incorporate the galactic bar \citep{Behzad2021, Behzad2022, Tahmasebzadeh2023}.
Despite the capability of deriving stellar orbit distributions through the orbit-superposition method, it is important to note that these distributions have limited accuracy, with certain fine-scale structures remaining undetermined. In the case of the CALIFA sample, each of the 300 galaxies has been coarsely divided into dynamically cold, warm, hot, and counter-rotating (CR) components, with the luminosity fractions and morphological characteristics of each component quantified \citep{zhu2018c}. The precision and uncertainty of these results have been thoroughly tested and understood. The dynamically cold and hot+CR components can be roughly linked to the photometrically defined thin disk and bulge, respectively. The warm component could be a mixture of thick disk and pseudo-bulge \citep{zhu2018c}. The CR component sometimes constructs a CR disk, but such cases are rare and are not statistically important in the CALIFA sample.

In this study, our objective is to bridge the gap between observations and simulations. To achieve this, we adopt a consistent methodology to define cold, warm, hot, and CR components for both TNG50 and TNG100 galaxies of IllustrisTNG suites, mirroring the approach used for CALIFA galaxies. We then establish a benchmark for the simulations by directly comparing the luminosity fractions and morphological properties of these components with those observed in CALIFA galaxies. Subsequently, we trace the formation of these components, with a particular emphasis on understanding the nature of the dynamically hot bulge component in TNG50 galaxies across the Hubble sequence and within a broad mass range. 
 
The paper is organized in the following way: we introduce the observations and simulations in Section~\ref{sec:samples}, describe the method in Section~\ref{sec:Methodology}, compare the TNG50, TNG100, and CALIFA galaxies in Section~\ref{sec:comparison}, quantify the formation of different structures in TNG50 in Section~\ref{sec:origins}. We discuss our results in Section~\ref{sec:discuss} and conclude in Section~\ref{sec:conclusion}.

\section{Observations and simulations}
\label{sec:samples}

The CALIFA survey encompasses galaxies of all morphological types across the Hubble diagram, exhibiting a diverse range of kinematic properties, from dispersion-dominated elliptical galaxies to rotation-dominated disk galaxies \citep{sanchez2012, falcon2017}. Notably, all CALIFA galaxies are situated in the field, and the CALIFA mother sample is considered to be representative of galaxies in the nearby universe within the stellar mass range of 10$^{8.7}$ \Msun to 10$^{11.9}$ \Msun. More than 95$\%$ of the galaxies in the mother sample are in the mass ranges of 10$^{9.7}$ \Msun to 10$^{11.4}$ \Msun where the selection function is well defined \citep{walcher2014califa}.

The CALIFA Data Release 2 (DR2) has made available data for 300 galaxies selected from the CALIFA mother sample, with 260 of them being isolated and devoid of kinematic biases introduced by dust lanes. A uniform application of the triaxial orbit-superposition Schwarzschild method \citep[e.g.][]{schwarzschild1979, van2008} is employed for each of the 260 galaxies \citep{zhu2018nature}. This method yielded the stellar orbit distribution within the effective radius $R_e$, presented as a probability density distribution of orbits within $R_e$ with respect to circularity $\lambda_z$. Based on the model, each galaxy is dynamically decomposed into four components: the cold ($\lambda_z> 0.8$), warm ($0.25<\lambda_z<0.8$), hot ($|\lambda_z| < 0.25$), and counter-rotating ($\lambda_z < -0.25$) component. The luminosity fraction of each of these components as a function of stellar mass is determined.

The result represents an observationally determined orbit distribution for galaxies in the nearby universe after sample corrections are applied to ensure a volume-average representation. To quantify morphological properties, the 3D density distribution of each component, constructed using orbits from the model, is projected onto the observational plane with an edge-on view \citep{zhu2018c}. 
The combination of luminosity fractions and morphological properties of these four dynamical components provides a comprehensive description of galaxy structures, enabling direct comparisons with many galaxies in cosmological simulations.

\subsection{IllustrisTNG simulations and TNG50}
\label{subsec:TNG50}

Among the various cosmological hydrodynamical simulations on galaxy formation and evolution, IllustrisTNG simulations \citep[hereafter TNG,][]{Springel2018, Marinacci2018, Naiman2018, Pillepich2018b, Nelson2018} have emerged as a major success in replicating a wide range of observational findings \citep{Nelson2019release}. These accomplishments encompass several key areas, including the galaxy mass-size relation at $0<z<2$ \citep{Genel2018}, the dimensions and heights of gaseous and stellar disks \citep{Pillepich2019}, galaxy colors, the stellar mass-age/metallicity relationship at $z\sim0$, which are consistent with SDSS results \citep{Nelson2018}, resolved star formation in star-forming galaxies \citep{ENelson2021}, and the ability to accurately model the characteristics of stellar orbit distributions in the CALIFA survey \citep{Xu2019} and the kinematics of early-type galaxies, in line with ATLAS-3D, MaNGA, and SAMI data \citep{Pulsoni2020}. 

We refer the reader to the methodological papers by \citealt{Weinberger2017, Pillepich2018a} for a complete understanding of the numerical and physical models implemented in the TNG simulations. 
The suite comprises three flagship simulations: TNG50, TNG100, and TNG300, each corresponding to different cosmological volumes of $(50\,\rm Mpc)^3$, $(100\,\rm Mpc)^3$, and $(300\,\rm Mpc)^3$, with associated stellar particle resolutions of $8.5 \times 10^{4}$\ \Msun, $1.4 \times 10^{6}$\ \Msun, and $1.1 \times 10^{7}$\ \Msun, respectively \citep{Nelson2019release}.

The luminosity fractions of the cold, warm, and hot components in TNG100 galaxies closely align with those of CALIFA galaxies within the stellar mass range of $M_*=10^{9.7}-10^{11.4} \,$\Msun\, \citep{Xu2019}. However, the enhanced resolution of TNG50 significantly improves its capacity to resolve galactic structures, particularly within the inner 1 kpc region and the thin disk.

\subsection{Galaxy samples selected from TNG100 and TNG50}
\label{subsec:Sample}
    \begin{figure*}
    \centering
    \includegraphics[width=2.00\columnwidth]{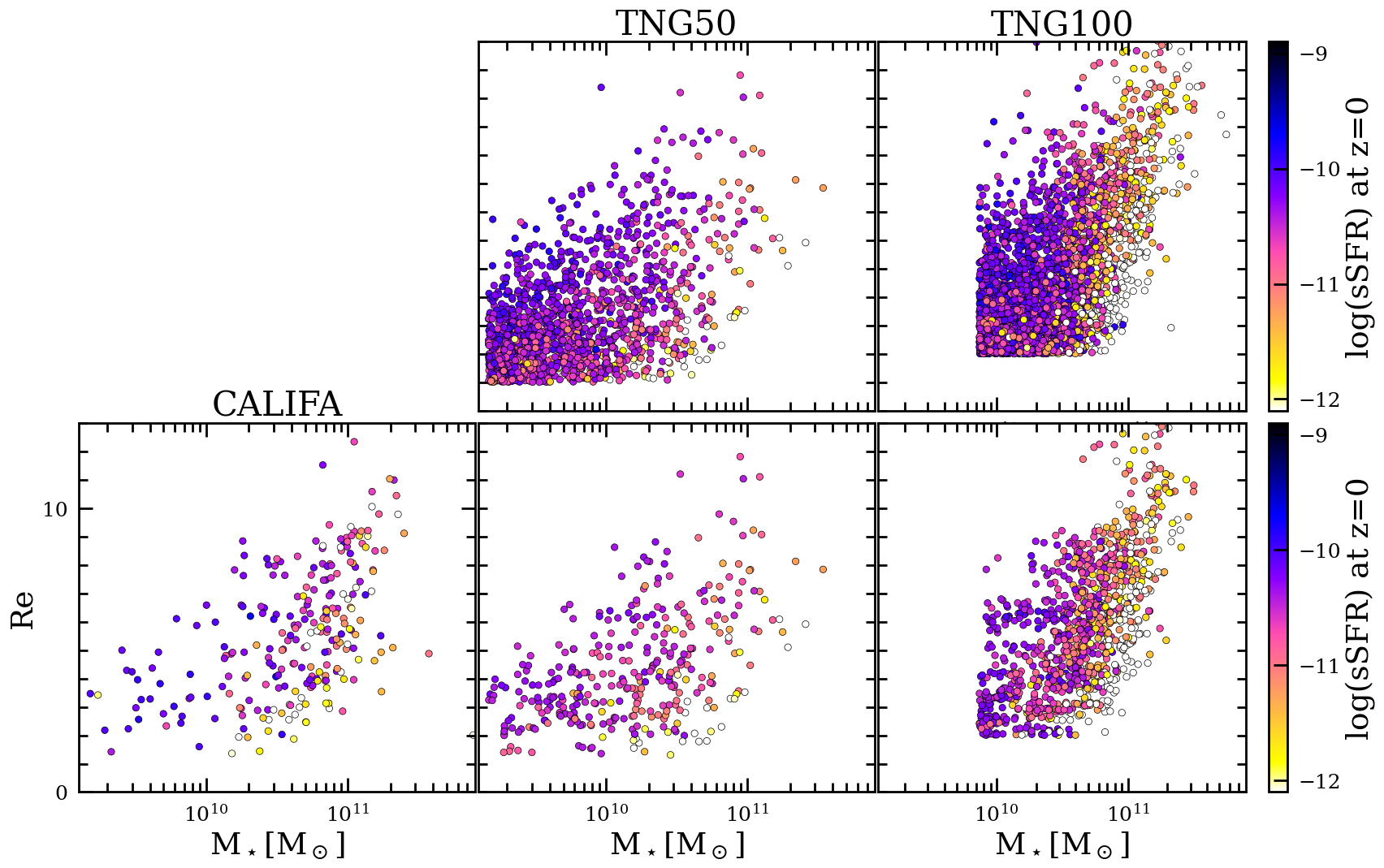}
        \caption{Stellar mass vs. $R_e$ for CALIFA (left lower), TNG50 (middle) and TNG100 (right) sample. The galaxies are color-coded by their sSFR, as indicated by the colorbar. The upper panels show the whole sample of central galaxies in the TNG50 and TNG100 selected from the mass-size plane. The bottom panels show samples that matched the CALIFA in mass, size, and sSFR.}
        \label{fig:sample}
    \end{figure*}

We selected simulated galaxies based on their galaxy stellar mass, and the comparison between observations and simulations mainly concentrates on stars within one effective radius $R_e$ of galaxies, aligning with the typical data coverage of most CALIFA galaxies. The $R_e$ of CALIFA galaxies is defined as the half-light semi-major axis measured from the r-band images. For the TNG galaxies, we define galaxy stellar mass as the total mass of all stars that are gravitationally bound to a galaxy according to the Subfind algorithm.
We adopt a similar approach as that of CALIFA galaxies to calculate $R_e$ by first randomly projecting the particles to the 2D observational plane and then selecting the 2D elliptical radius that covers half of the total r-band luminosity.

To mirror the characteristics of the CALIFA sample in \citet{zhu2018nature}, our selection criteria encompass galaxies at $z=0$ (snapshot 99) that
{\bf (1)} are central galaxies;
{\bf (2)} are not in the midst of a merger event;
{\bf (3)} possess stellar masses within the range of $M_*$ $\in$ [$10^{9}$,$10^{12}$] \Msun\, for TNG50 and $M_*$ $\in$ [$10^{9.7}$,$10^{12}$] \Msun\, for TNG100;
{\bf (4)} $R_e>1$ kpc for TNG50 and $R_e>2$ kpc for TNG100.
We include lower-mass galaxies from TNG50 with their structures well-resolved with higher mass and spatial resolution. The TNG50 sample we select covers the mass range of the full sample of the CALIFA galaxies.
Consequently, our analysis encompasses 1838 galaxies from TNG50 and 5356 galaxies from TNG100, spanning various morphological types.
For a fair comparison, we further compose a sample from TNG50 and TNG100 to match the CALIFA galaxies in $M_*$, $R_e$, and specific star formation rate (sSFR). We take the three TNG50 counterparts and five TNG100 counterparts for each CALIFA galaxy with the smallest differences in $M_*$, $R_e$ and sSFR (only galaxies with $M_* \ge 10^{9.7}$ \Msun\ are included for TNG100). This results in 281 TNG50 galaxies and 891 TNG100 galaxies selected out; a large fraction of them are selected more than once; we count them repeatedly to keep the whole sample representative of CALIFA.

In Fig. \ref{fig:sample}, we present the CALIFA, TNG50, and TNG100 samples in the space of stellar mass vs. $R_e$, colour-coded by the sSFR. The galaxies from CALIFA and TNG50 show broad consistency in the distribution of mass, size, and sSFR. The passive population in TNG100 with sSFR $< 12.5$ does not exist in the CALIFA sample. There are a limited number of galaxies with stellar masses $M_* \ge 10^{11}$\, \Msun\, in TNG50, which potentially might introduce some statistical biases.

Within the framework of this study, we specifically compare galactic structures in the CALIFA, TNG100, and TNG50 galaxies, including the luminosity fractions and morphological properties of each orbital component. In addition, we traced the formation of these structures using the TNG50 simulation.

    \begin{figure}
    \centering
    \includegraphics[width=\columnwidth]{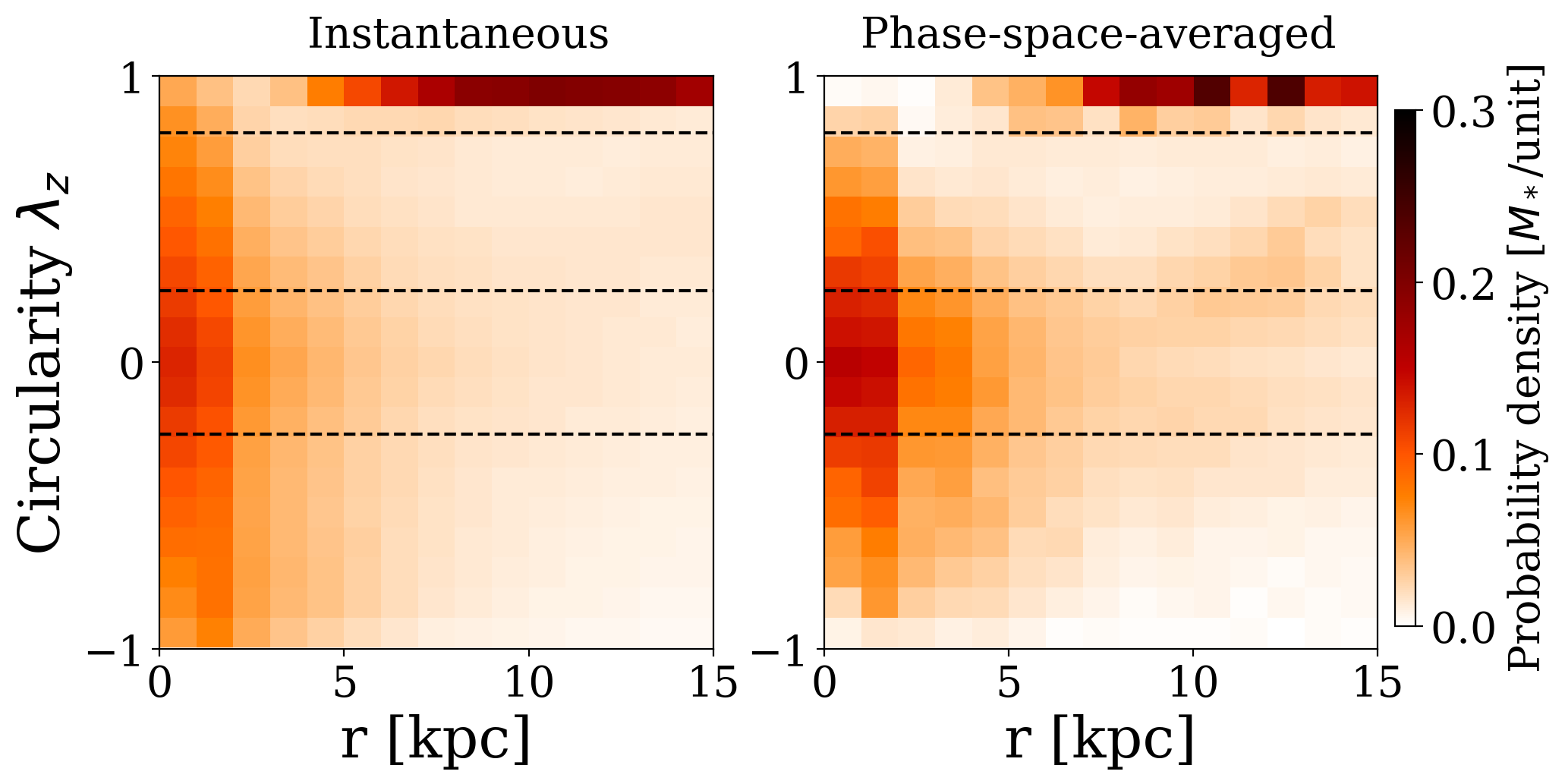}
        \caption{The stellar orbit distribution of a TNG50 galaxy represented as the probability density distribution in the phase space of circularity $\lambda_z$ vs. radius $r$. On the left, we use the $\lambda_z$ and $r$ calculated from instantaneous positions and velocities of the stellar particles. On the right, we use the orbital $\lambda_z$ and $r$ obtained through a phase-space averaging approach, where particles with similar energy and angular momentum are grouped. Dashed lines are used to demarcate the boundaries separating the cold ($\lambda_z > 0.8$), warm ($0.25 < \lambda_z < 0.8$), hot ($|\lambda_z| < 0.25$), and counter-rotating (CR) ($\lambda_z < -0.25$) components based on circularity.}
        \label{fig:l_r}
    \end{figure}
    
    \begin{figure*}
    \centering
    \includegraphics[width=2.00\columnwidth]{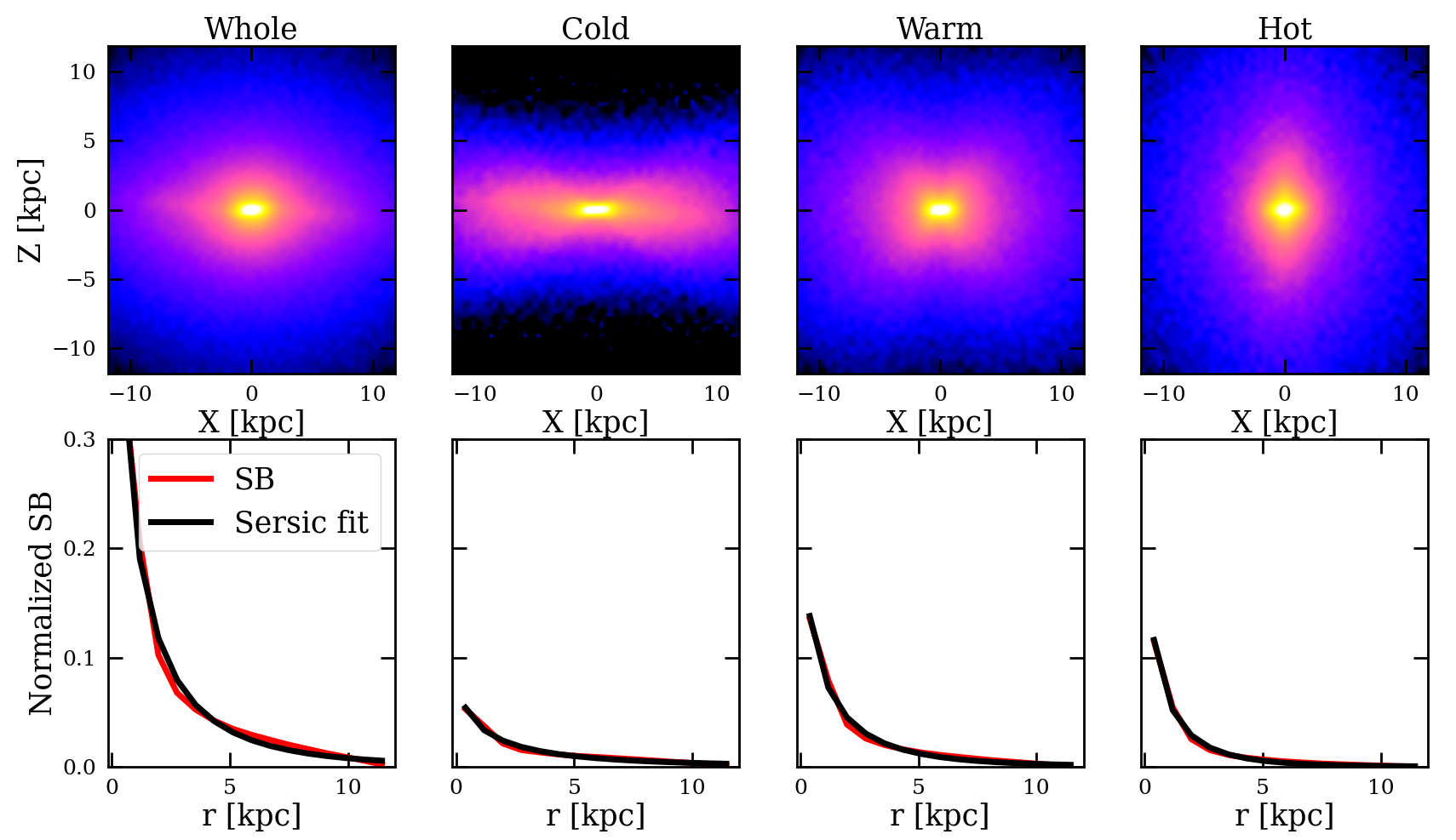}
        \caption{An illustrative example showcases the morphology of various orbital components. The panels from left to right represent the entire galaxy, followed by the cold, warm, and hot components, respectively. In the top panel, we observe the edge-on surface brightness, while the bottom panel exhibits the surface brightness along the major axis. The latter is given in red and is fitted by a S\'ersic profile (black).}
        \label{fig:sersicfit}
    \end{figure*}

\section{Methodology}
\label{sec:Methodology}
\subsection{Orbital structure decomposition}
\label{subsec:decomposition}
The stellar orbit distributions of CALIFA galaxies, as determined through the orbit-superposition method, are described as the probability density distribution of the stellar orbits in the phase space of the time-averaged radius $r$ versus the circularity $\lambda_z$. In this context, $\lambda_z$$\equiv$$J_z$/$J_{\rm max}$($E_b$) of an orbit represents the angular momentum around the short axis $J_z$ normalised by the maximum angular momentum achievable by a circular orbit with the same binding energy, denoted as $J_{\rm max}$($E_b$). Both $r$ and $J_z$ of an orbit are average values of a sample of particles sampled with an equal time step along the orbital trajectory.
1`
From a simulated galaxy, the 6D phase-space information of stellar particles provides us with instantaneous data at a given snapshot. However, it is crucial to note that particles moving in a specific orbit do not conserve their radius and also angular momentum in triaxial systems.
To directly compare with the output of the orbit-superposition model, we employ an approximate method to calculate $r$ and $\lambda_z$ for the orbits on which the stellar particles move. We achieve this through phase-space averaging, as follows: we categorise all particles within a simulated galaxy into bins within a 3D grid defined by angular momentum $J_z$, total angular momentum $J$, and bounding energy $E_b$. This categorisation is made under the assumption that particles occupying the same bin follow similar orbits, albeit in a different phase space. We then compute the average values of $r$ and $\lambda_z$ for the particles within each bin, considering these values as the orbital $r$ and $\lambda_z$ of these particles.

In Fig. \ref{fig:l_r}, we present the stellar orbit distribution of a TNG50 galaxy. The left panel shows the instantaneous $r$ versus circularity $\lambda_z$, while the right panel displays the phase-space-averaged $r$ versus circularity $\lambda_z$. In principle, box orbits exhibit circularity $\lambda_z \sim 0$, and particles sampled from a box orbit can encompass a wide range of instantaneous $\lambda_z$, extending even to $\lambda_z\sim 1$, Consequently, these particles may introduce contamination in the cold disk component if using the instantaneous $\lambda_z$. When considering the phase-space-averaged $\lambda_z$, the distribution of particles from box orbits narrows more closely around zero, although not precisely at $\lambda_z = 0$. This reduces the contamination of box orbit particles to the dynamically cold component.

Using the stellar orbit distribution based on phase-space averaged $\lambda_z$, we classify each galaxy into four distinct dynamical components: cold ($\lambda_z \ge 0.8$), warm ($0.25 \le \lambda_z \le 0.8$), hot ($|\lambda_z|\le 0.25$), and counter-rotating (CR) components ($\lambda_z \le -0.25$).  For a direct comparison with the CALIFA galaxies, we restrict our analysis to particles within $R_e$, which aligns with the spatial coverage of the CALIFA data.

The stellar orbit using phase-space-averaged $r$ and $\lambda_z$ is shown to work well in representing cold, warm, and hot components similar to those using time-averaged values \citep{Zhu2022a,zhu2018nature}. In addition, when evaluating the performance of the orbit superposition model against simulations, the ground truth of the simulations is described in the phase-space-averaged $r$ and $\lambda_z$, and the luminosity fractions of various components derived from the orbit superposition model are consistent with the ground truth \citep{zhu2018a, zhu2018nature}. Therefore, we think it is fair to compare the phase-space-averaged stellar orbit distributions of simulated galaxies with those of CALIFA galaxies obtained from the orbit superposition model.

The cold and hot+CR components are the main ingredients of the morphologically defined thin disk and bulge, respectively, while the warm component could be a mixture of thick disk and pseudo-bulge \citep{zhu2018c}. Our definition of cold, warm, and hot + CR components within 1 $R_e$ is statistically consistent with the cold disk, warm disk, and bulge identified by the automated Gaussian Mixture Models \citep[auto-GMM in][]{du2019identifying}, while the hot + CR component in the outer regions roughly corresponds to their halo. The bar or pseudobulge is not well separated in \citet{du2019identifying} either, and may be included mostly in the warm disk component.

\subsection{Description of morphology}
\label{subsec:morp}
In the case of CALIFA galaxies, the 3D density distribution of each component is reconstructed using particles sampled from the corresponding orbits; subsequently, these distributions are projected in an edge-on fashion onto the 2D observational plane. In line with this approach, we reconstruct the edge-on surface brightness of each component using the same method for the TNG galaxies, with one case shown in Fig. \ref{fig:sersicfit}.
We then describe the morphology of each component using two key parameters: intrinsic flattening $q_{R_e}$ and the S\'ersic index $n$, obtained precisely in the same way as defined for the CALIFA galaxies in \citet{zhu2018c}.

The $q_{R_e}$ of each component is calculated using only the stellar particles within $r<Re$: 
\begin{equation}
q_{R_e} = \sqrt{(\sum_i L_i  z_i^2) / (\sum_i L_i x_i^2)},
\end{equation}
where $L_i$ is the r-band luminosity, $x_i$ and $z_i$ are particle positions along the long and short axis of that component, respectively.

We fit the S\'ersic profile:
\begin{equation}
    \Sigma(r)=\Sigma_e 10^{-b_n((r/r_s)^{1/n}-1)}
	\label{eq:sersiceq}
\end{equation}
to the projected edge-on surface brightness (SB) along the major axis of each component, with two free parameters, the effective surface density $\Sigma_e$ and S\'ersic index $n$.
As shown in Fig. \ref{fig:sersicfit}, our S\'ersic profile fittings align closely with the simulated profile, demonstrating a good match.

\subsection{\textit{In-situ} and \textit{ex-situ} stars}
In the TNG simulations, the identification of halos, subhalos, and fundamental galaxy properties is based on the Friends-of-Friends (FoF) and \textsc{Subfind} algorithms \citep{Springel2001, Dolag2009}. 

To comprehensively trace the histories of galaxies in the simulations, we employ merger trees constructed by the \textsc{Sublink-gal} code, primarily based on the baryonic components of subhaloes \citep{Rodriguez2015}. In this algorithm, each galaxy is assigned a unique descendant. It allows us to identify, for each galaxy, its "mergers" – that is, secondary galaxies with well-defined "infall" times that eventually merge with it and do not exist as individually identified \textsc{Subfind} objects after the "coalescence." Additionally, for any given galaxy at a specific time, we can also identify its "satellites," which are galaxies with well-defined "infall" times or orbits that orbit around it, potentially merging in the future.

Thanks to the merger trees, we can distinguish between two categories of stars: \textit{in-situ} and \textit{ex-situ} stars \citep{Rodriguez-Gomez2019, Pillepich2018a}. Stars in a galaxy at $z=0$ are classified as \textit{in-situ} if they originated from a progenitor belonging to the galaxy's primary progenitor branch. In contrast, stars are labeled \textit{ex-situ} if they are formed outside the galaxy and subsequently accreted. \textit{Ex-situ} stars are predominantly composed of stellar particles stripped from satellites and accreted into the galaxy. These stars are primarily the result of mergers with other objects that have already merged at the time of examination. However, a portion of \textit{ex-situ} stars may also arise from the stripping of orbiting satellites that have not yet been destroyed.

Throughout the paper, we define the merger mass ratio $M_{*, \rm sat}/M_{*, \rm main,z=0}$ by the stellar mass of the satellite at its peak over the stellar mass of the main progenitor at $z=0$, and we define mergers with mass ratios of $>1:3$, $1:3-1:10$, and $<1:10$ as major, minor, and mini mergers, respectively. This suggests that significant mergers take place at a later stage, resulting in the influx of a substantial number of {\it ex-situ} stars. These mergers strongly influence the pre-existing disks, while earlier mergers, which occur at a formative stage, have minimal impact on the disks that develop subsequently.

\subsection{The circularity of \textit{in-situ} stars at birth}
In the TNG50 simulation, we can determine the circularity of \textit{in-situ} stars at the time of their formation. To achieve this, we rely on the 'ParticleIDs' catalog, which provides a unique ID for each star or wind particle and remains constant throughout the simulation. For each \textit{in-situ} stellar particle, we backtrack its formation by using the 'ParticleIDs.' We pinpoint the snapshot in which it first appears as a star and subsequently calculate its phase-space-averaged circularity $\lambda_z$ at this specific snapshot, considering this value as the particle's orbital circularity at birth.

Note that we did not manage to identify the birth snapshot for all \textit{in-situ} particles. This inability to pinpoint the exact birth snapshot can be attributed to the intricacies of the \textsc{Subfind} algorithm. However, the results are not significantly affected by this limitation, as the number of particles for which we cannot determine the birth snapshot is not more than $1\%$ of the total \textit{in-situ} stars.

\section{Comparison between observations and simulations}
\label{sec:comparison}
\subsection{Luminosity fractions of four orbital components}
\label{subsec:flum}

    \begin{figure*}
    \centering
    \includegraphics[width=2.00\columnwidth]{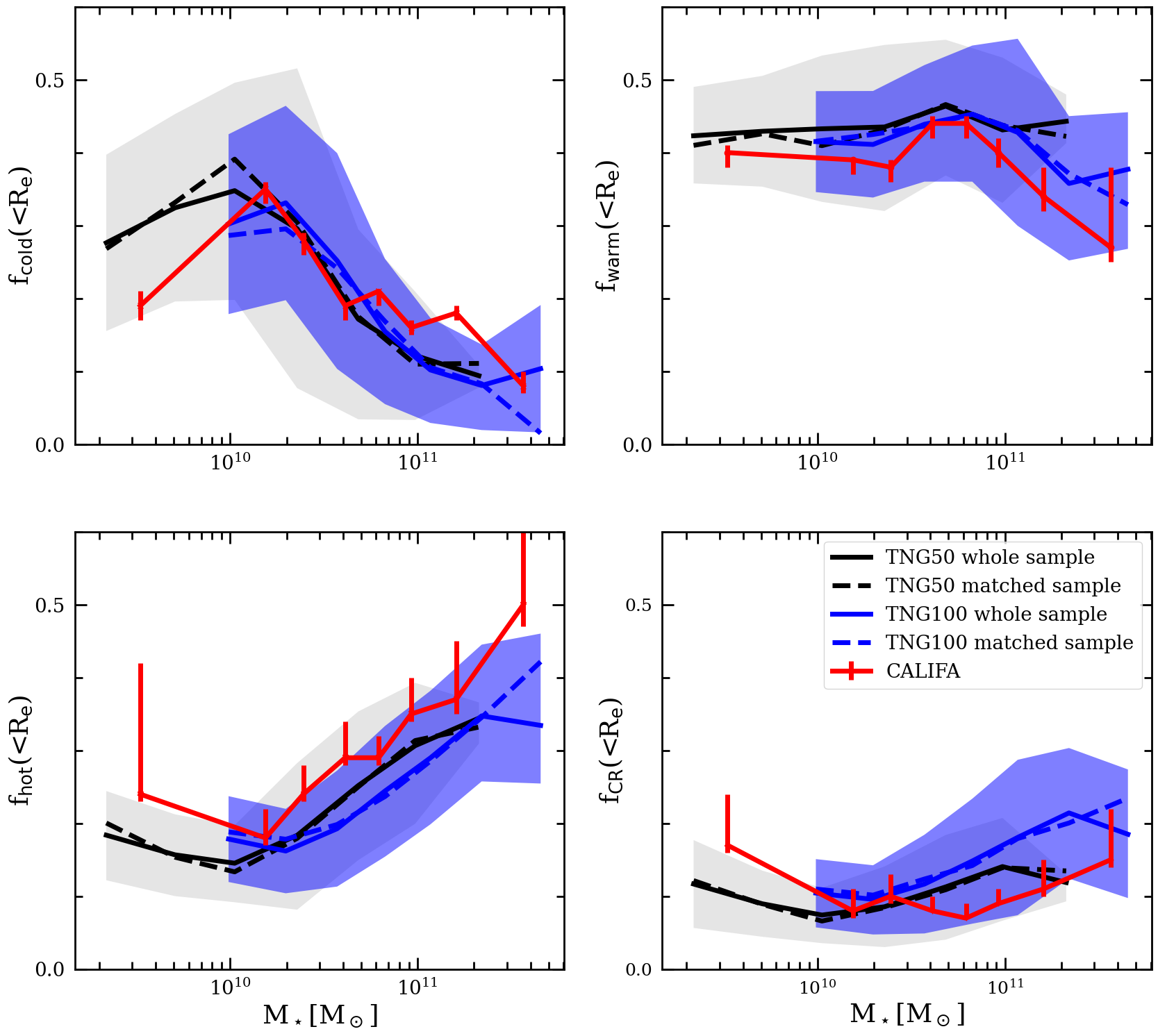}
        \caption{Comparison of luminosity fractions of four orbital components among CALIFA, TNG50, and TNG100 galaxies at $z\sim 0$. The trend across the four panels each represents the luminosity fraction of the cold, warm, hot, and CR components as functions of stellar mass. Each red solid curves represent the CALIFA galaxies from \citet{zhu2018nature}, and the associated error bars represent the $1 \sigma$ uncertainties. The black and blue curves represent the mean values of luminosity fraction as a function of stellar mass for TNG50 and TNG100 galaxies, respectively. The solid lines are for the whole sample and the dashed lines for the sample matched CALIFA in mass, size, and sSFR. The shaded areas represent the corresponding 1$\sigma$ scatters, indicating that $68\%$ of the galaxies fall within these regions. Both TNG50 and TNG100 broadly replicate the luminosity fractions of the four components and their dependence on stellar mass, as observed in the CALIFA galaxies.}
    \label{fig:Flum}
    \end{figure*}

With the orbital decomposition, we can directly compare the internal orbital structures of observed galaxies to their simulated counterparts. In Fig. \ref{fig:Flum}, we show the luminosity fractions of the cold, warm, hot, and CR components as functions of stellar mass for the TNG50, TNG100, and CALIFA galaxies.
The luminosity fractions of different orbital components from CALIFA galaxies are also consistent with those obtained from SAMI \citep{Santucci2022} and MaNGA \citep{Jin2020} galaxies that we do not show here. For TNG50 and TNG100 galaxies, we show the results from the whole sample and the sample rigorously matched to CALIFA galaxies in mass, size, and sSFR: there is no noticeable difference between the two samples.

The luminosity fractions of the four components and their dependence on stellar mass exhibit broad consistency among TNG50, TNG100, and CALIFA, particularly within the intermediate mass range of $M_*$ $\in$ [$10^{10}$,$10^{11}$] \Msun. 
In both TNG and CALIFA galaxies, the luminosity fraction of the cold component peaks at $M_*\sim 1-2 \times 10^{10}$ \Msun, and decreases at lower and higher mass ranges, with the fractions of the hot and CR components increasing. However, the luminosity fraction of the warm component remains largely unaffected, remaining relatively constant in galaxies with stellar masses $M_*\lesssim 10^{11}$ \Msun\ but declining in the most massive galaxies. The consensus among the three datasets suggests that: (1) the TNG model effectively replicates the kinematics of actual galaxies; and (2) the kinematics portrayed by the TNG model remain consistent across varying resolutions, indicating resilience to resolution effects.

However, there are some noticeable differences between CALIFA and TNG galaxies at the low and high ends of the mass range. In galaxies with $M_*\le 10^{10}$ \Msun, the luminosity fraction of the cold component ($f_{\rm cold}$) is slightly higher in TNG50 than in CALIFA (0.25 in TNG50 vs. 0.19 in CALIFA), while $f_{\rm hot}$ and $f_{\rm CR}$ are slightly lower. It is important to note that the CALIFA sample is not complete and the data quality is relatively low at $M_* < 10^{9.7}$ \Msun, due to the limited spectral resolution, which could contribute to this offset.

At the high mass end, with $M_*\ge 2\times 10^{11}$ \Msun, TNG50 and TNG100 galaxies exhibit lower fractions of the hot component and higher fractions of the warm components compared to CALIFA galaxies ($f_{\rm hot}$: 0.34 in TNG vs. 0.50 in CALIFA; $f_{\rm warm}$: 0.43/0.37 in TNG50/100 vs. 0.27 in CALIFA). This discrepancy might be partly caused by the approximations involved in obtaining circularity $\lambda_z$ by phase-space averaging, as described in Section~\ref{sec:Methodology}. Massive elliptical galaxies are dominated by box orbits with $\lambda_z\sim 0$, and their instantaneous $\lambda_z$ have a wide distribution. Phase-space averaging can somewhat narrow the $\lambda_z$ distribution, but it may still categorise some particles on dynamically hot orbits as warm or CR components due to method imperfections. These biases in classifying hot, warm, and CR orbits may be present in all galaxies but are most pronounced in the most massive galaxies dominated by hot orbits.

The variation of orbit components shown above is consistent with the findings reported in \citet{liang2024connection} that galaxies with MW mass ($M_*\sim 10^{10.5}$ \Msun) have the thinnest disks settled. This marks the point where TNG50 galaxies undergo, on average, a morphological transformation from irregular or clumpy disk-like systems to symmetric Hubble-type systems on the main sequence at $z \sim 1$  \citep{varma2022building}. This transformation results in an increase in density within the central region of 1 kpc, mainly composed of hot orbits due to the central truncation of the dynamically cold disk \citep{du2020kinematic}. 
\subsection{Morphology of 4 orbital components}
\label{subsec:morph}
    \begin{figure*}
    \centering
    \includegraphics[width=2.00\columnwidth]{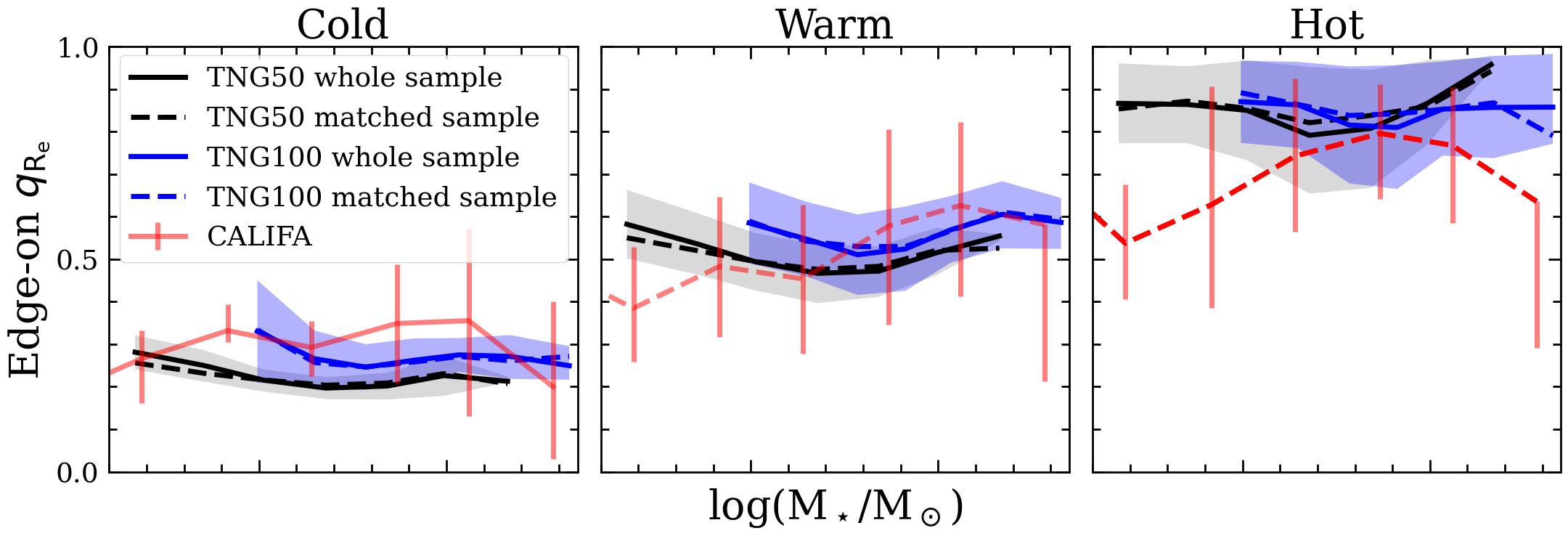}
        \caption{The intrinsic flattening ($q_{\rm R_e}$) of each component as functions of the galaxy's stellar mass $M_*$. The three columns, from left to right, represent the cold, warm, and hot components, respectively. Line styles and colors are same as \ref{fig:Flum}.}
    \label{fig:qe}
    \end{figure*}

    \begin{figure*}
    \centering
    \includegraphics[width=2.00\columnwidth]{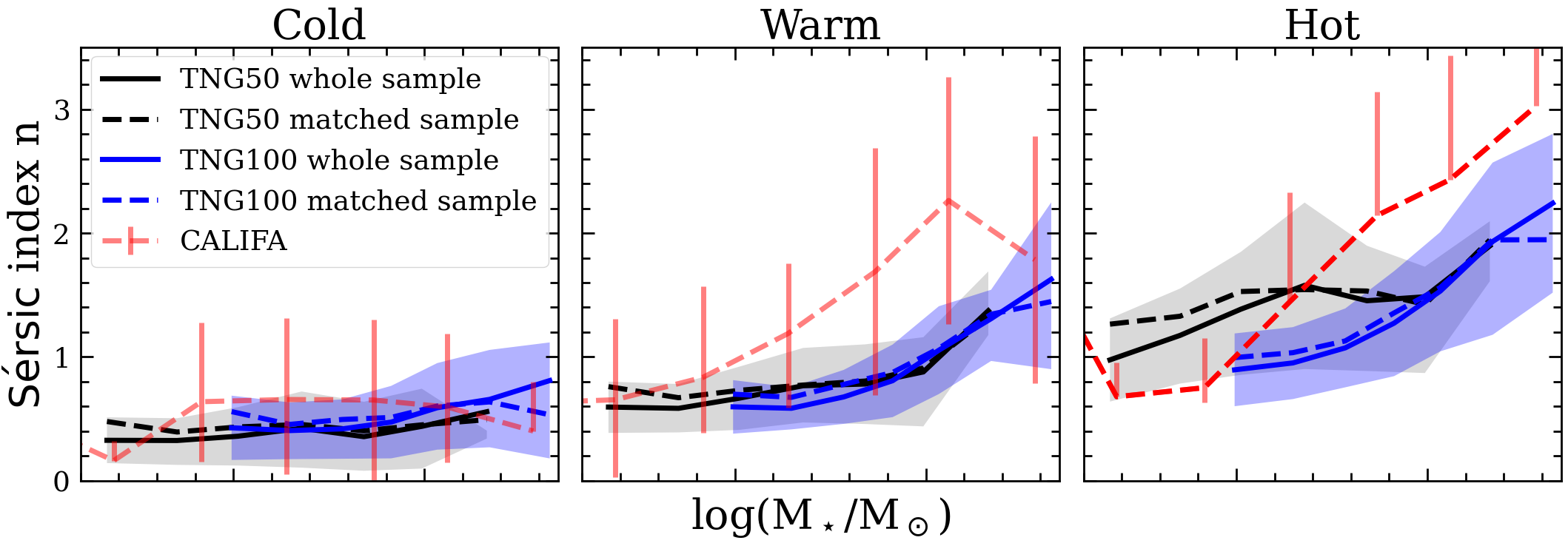}
        \caption{Same as Fig. \ref{fig:qe}, but for the S\'ersic index $n$.}
    \label{fig:sersic}
    \end{figure*}

We further compare the morphology of the orbital components between the TNG50, TNG100, and CALIFA galaxies. As described in Section~\ref{sec:Methodology}, we define two parameters to describe the morphology: the intrinsic flattening $q_{\rm Re}$ calculated within $R_e$ and the S\'ersic index $n$. Both parameters were calculated for the TNG galaxies exactly the same way as for the CALIFA galaxies \citep{zhu2018c}.

\subsubsection{Intrinsic flattening}
\label{subsub:qe}

In Fig. \ref{fig:qe}, we show the intrinsic flattening $q_{R_e}$ of the cold, warm, and hot+CR components as functions of stellar mass for TNG50, TNG100, and CALIFA galaxies. Generally, the dynamically hotter components appear to be rounder, as indicated by higher values of $q_{R_e}$, compared to the cold component. This trend is observed in both CALIFA and TNG galaxies, and it is expected from the correlation of galactic dynamics and morphology \citep{zhu2018nature}. For TNG50 and TNG100, we show the results of the whole sample and the matched sample, respectively. The results of the two samples are almost identical.

In CALIFA, the intrinsic flattening $q_{R_e}$ of the cold and warm components does not show a significant dependence on the galaxy's stellar mass, with median values of approximately 0.3 and 0.5, respectively. However, the hot component appears to be rounder in more massive galaxies, with $q_{R_e}$ increasing from $\sim 0.6$ for galaxies with stellar masses around $M_* \sim 10^9-10^{10}$ \Msun \, to $\sim 0.8$ with $M_*\sim 10^{11}$ \Msun.

For the cold and warm components, both TNG50 and TNG100 galaxies exhibit good consistency with the CALIFA galaxies in terms of their flattening characteristics and show no significant dependence on galaxy mass. However, the hot components in TNG50 and TNG100 have $q_{R_e} \sim 0.9$ with no apparent dependence on the galaxy mass. This suggests that the hot components in the TNG simulations are generally rounder than the hot components in CALIFA, particularly in galaxies with stellar masses below $M_*<10^{10}$ \Msun.

\subsubsection{S\'ersic Index}
\label{subsub:sersic}

In Fig. \ref{fig:sersic}, we show the S\'ersic index $n$ of the cold, warm, and hot+CR components as functions of the galaxies' stellar mass for the TNG50, TNG100, and CALIFA galaxies. 

In CALIFA galaxies, the cold component has a relatively constant S\'ersic index of $n\sim0.5$, throughout the stellar mass range. However, the warm and hot components are more spatially concentrated in more massive galaxies.
For galaxies with $M_* \sim 10^9-10^{10}$\ \Msun, $n$ is $\sim 1$ for the warm and hot components, while increasing to $\sim 2$ and $\sim 3$, separately, for galaxies with $M_*\sim 10^{11}$\ \Msun. Still, the samples matched to CALIFA galaxies show a similar trend to the whole sample of galaxies, for both TNG50 and TNG100.

In TNG50 and TNG100, the cold component shows good consistency with CALIFA galaxies, with similar S\'ersic index values and no apparent dependence on stellar mass. On the other hand, both the warm and the hot components exhibit a similar trend of increasing $n$ with galaxy stellar mass, which is consistent with CALIFA galaxies. However, the values of $n$ for the warm and hot components in the TNG simulations are smaller than those in CALIFA galaxies. The warm and hot components in the TNG galaxies are not as centrally concentrated as those in the CALIFA galaxies, consistent with the conclusion of \citet{du2020kinematic}.

Despite the morphology of CALIFA galaxies not being perfectly consistent with TNG simulations, the characteristics of the four kinematic components of observed galaxies are broadly reproduced in the simulations.

    \begin{figure*}
    \centering
    \includegraphics[width=1.8\columnwidth]{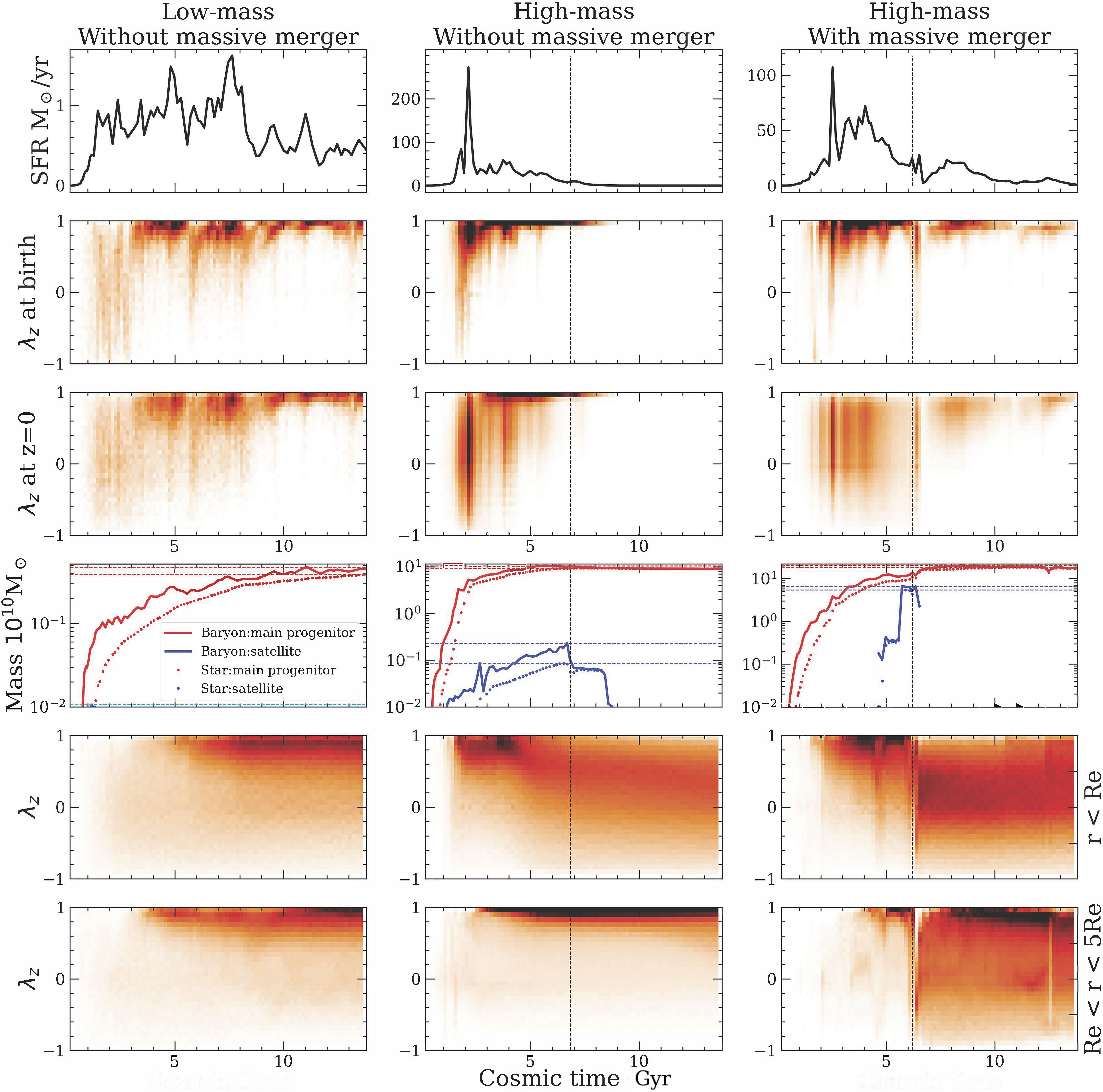}
        \caption{The evolution of galaxy structures in three typical cases from left to right. For each galaxy, we present the star formation history and its kinematics in the top three rows:
        {\bf (1)} star formation history: the star formation rate as a function of cosmic time; {\bf (2)} the birth circularity distribution within $5R_e$ for stars formed at different cosmic epochs; {\bf (3)} the present day ($z=0$) circularity distribution within $5R_e$ for stars born at different cosmic epochs.
        In the subsequent three rows, we show the mass assembly history of the main progenitor; {\bf (4)} the red solid and dotted curves represent the baryonic and stellar mass evolution of the main progenitor over cosmic time, and the blue lines represent those of secondary galaxies that merged with the main progenitor. The vertical dashed lines indicate the time of mergers, if any; {\bf (5)} circularity distribution of stars at $r<R_e$ of the main progenitor at different cosmic epochs; {\bf (6)} circularity distribution of stars at $R_e<r<5R_e$ of the main progenitor.
        \textbf{Left}:
        In the left panel, we show a low mass galaxy ($M_* = 10^{9.76}$ \Msun) that has not experienced any major mergers. Stars in this galaxy are formed in a relatively warm state, and their circularity distribution at birth remains largely unchanged until the present day. There is no significant change in the stellar circularity distribution in the main progenitor over time.
        \textbf{Middle}:
        In the middle panel, we show a high mass galaxy ($M_* = 10^{11.12}$ \Msun) that also lacks major mergers, with the most significant merger having a mass ratio $M_{*, \rm sat}/M_{*, \rm main,z=0} \le 0.01$). The majority of stars in this galaxy are born in a dynamically cold state. However, stars born during the star formation peak at $t\sim2$ Gyr and become dynamically warm/hot by $z=0$. The circularity distribution of stars in the main progenitor within $r<R_e$ gradually transitions from dynamically cold to dynamically hot, while the stars at $R_e<r<5R_e$ remain dynamically cold. 
        \textbf{Right}: 
        In the right panel, we show a high mass galaxy ($M_* = 10^{11.41}$ \Msun) that experienced a major merger with ($M_{*,\rm sat}/M_{*,\rm main, z=0} \sim 0.2$). Most stars in this galaxy are born in a cold state, and stars born before the merger event are warm/hot at $z=0$. The circularity distribution of stars in the main progenitor, both in the central and outer regions, has undergone dramatic changes due to the merger.}
    \label{fig:merger}
    \end{figure*}

    \begin{figure*}
    \centering
    \includegraphics[width=1.7\columnwidth]{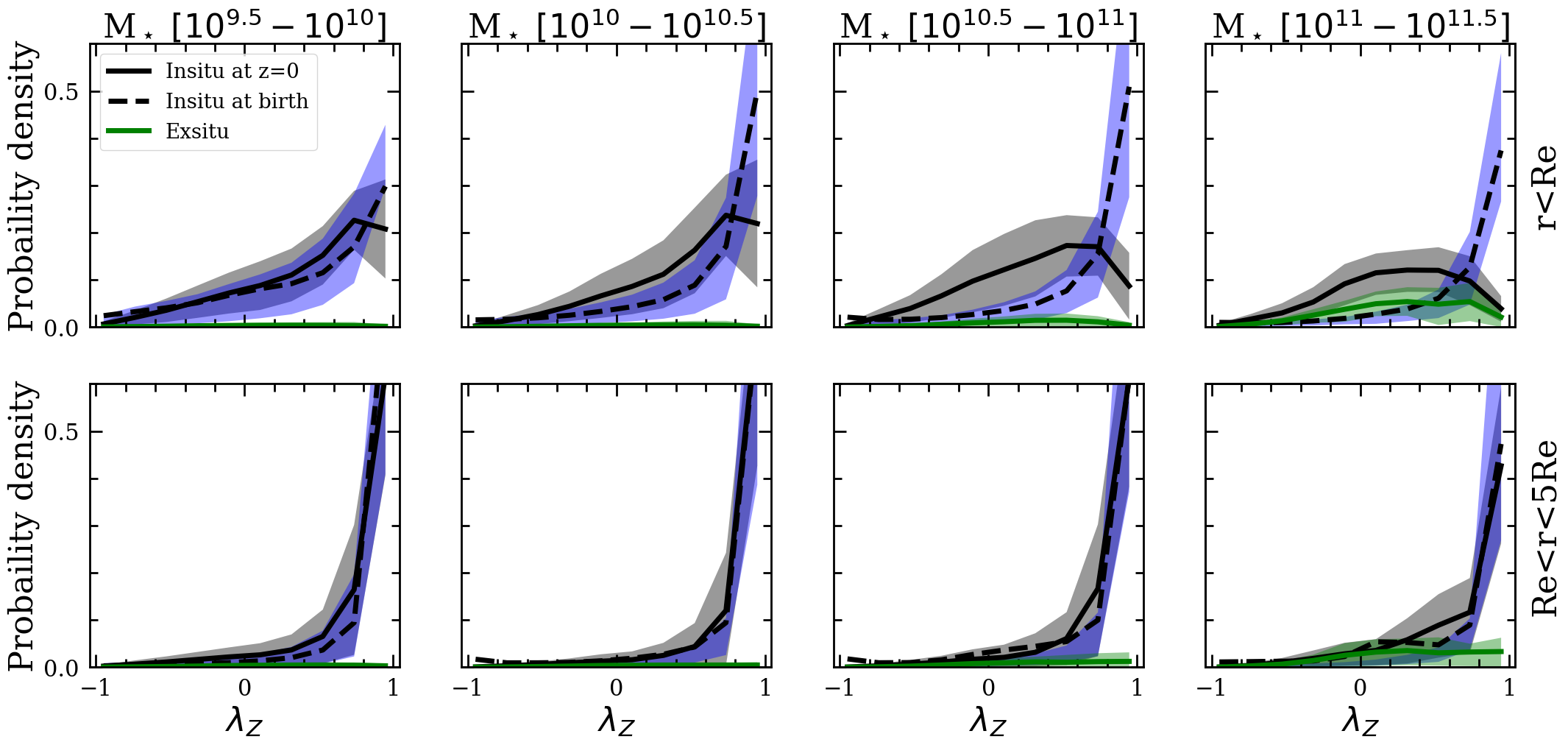}
        \caption{The circularity distribution of \textit{in-situ} and \textit{ex-situ} stars at $z=0$ in TNG50, comparing to the circularity at birth for the \textit{in-situ} stars. The columns from left to right represent galaxies with increasing stellar masses, and the top and bottom rows represent stellar particles in the inner $R_e$ and in the outer regions $R_e<r<5R_e$, respectively. Within each panel, the solid black and green curve represent the present-day circularity distribution at $z=0$ for \textit{in-situ} and \textit{ex-situ} stars, respectively, while the dashed black curve represents the birth circularity distribution of \textit{in-situ} stars. The shaded areas indicate the 1$\sigma$ scatters among galaxies in each bin. Combining \textit{in-situ} and \textit{ex-situ} stars results in the overall present-day circularity distribution of stars in galaxies at $z=0$.
        In lower mass galaxies, the \textit{in-situ} stars at $z=0$ maintain a circularity distribution similar to their birth conditions, with little or no significant heating across the whole galaxy. In higher mass galaxies, the \textit{in-situ} stars are mostly born on cold orbits with peaking $\lambda_z \sim 1$. The stars in the inner $R_e$ have much lower $\lambda_z$ at $z=0$ compared to their condition at birth, indicating that significant heating occurred in the inner regions. The stars in the outer regions $R_e<r<5R_e$ have circularity distributions at $z=0$ similar to those at birth.}
    \label{fig:bieRe}
    \end{figure*}

\section{The physical origins of different components}
\label{sec:origins}
Given the consistency of internal structures of CALIFA and TNG50 galaxies as shown in the last section. Here we trace the formation of different orbital components, especially the dynamically hot+CR bulge component, using the TNG50 simulation.

\subsection{Illustration of three typical cases}
\label{subsub:cases}

In Fig. \ref{fig:merger}, we illustrate three typical pathways of galaxy structure evolution for galaxies with different stellar masses and merger histories. As described before, we define the merger mass ratio $M_{*, \rm sat}/M_{*, \rm main,z=0}$ by the stellar mass of the satellite at its peak over the stellar mass of the main progenitor at $z=0$, and we define mergers with mass ratios of $>1:3$, $1:3-1:10$, and $<1:10$ as major, minor, and mini mergers, respectively.

In the left column, we show a low-mass galaxy $M_*=5.7\times 10^9 $\ \Msun\, with no major mergers.
This galaxy, characterised by a gas-rich environment, exhibits a stable star formation rate over cosmic time. Star born in different periods were in different dynamical status: {\bf (1)} stars born on predominantly hot orbits in the early universe ($t\lesssim 3$ Gyr), displaying a wide distribution of $\lambda_z$ at birth; {\bf (2)} stars born in colder orbits at 3 Gyr $\lesssim$ t $\lesssim$ 8 Gyr, but still with some stars born on warm and hot orbits; {\bf (3)} only in the recent universe ($t \gtrsim 8$ Gyr), stars that are overwhelmingly born on cold orbits. 
The present-day $\lambda_z$ distribution of the stars at $z=0$ generally conserves their $\lambda_z$ distribution at birth. The $\lambda_z$ distribution of the main progenitor remains largely unchanged. As a result, the dynamically hot bulge in this galaxy at $z=0$ was born hot mostly in the early and intermediate universe.

In the middle column, we show a galaxy with a higher mass $M_*=1.3 \times 10^{11}$\ \Msun\, and still without major mergers, the most massive satellite merged contributing $\le$0.01 of the current stellar mass of the galaxy.
This galaxy had a starburst in the early universe ($t\sim 2$ Gyr), with most stars born on cold orbits and a small fraction of stars born in warm/hot orbits; the star formation rate drops rapidly after that and it keeps star forming for a long term ($2\lesssim t \lesssim 8$) with stars predominantly formed in cold orbits; the star formation quenched at $t\sim 9$ Gyr in this galaxy.
Despite the absence of major mergers, stars in the inner regions ($r<R_e$) of this galaxy are gradually heated at the epoch of $t\sim 5-7$ Gyr, forming a dynamically warm/hot bulge. No significant mergers or external disturbances were detected during the period. The stars that have been heated are spatially concentrated in the inner regions and mostly formed in the early starburst. Stars in the outer regions ($R_e<r<5R_e$), also formed later, maintain a dynamically cold disk until $z=0$.

In the right column, we show a massive galaxy ($M_*=2.6 \times 10^{11}$\ \Msun) that experienced a major merger with a mass ratio of $1:3$ at cosmic time $t\sim 6$ Gyr.
Before the major merger, most stars in this galaxy formed in dynamically cold orbits. The merger dramatically reshaped the galaxy, heating all stars formed before the merger. A sudden transition in the circularity distribution occurs during the merger, affecting stars in all regions. After the merger, stars in the inner region $r<R_e$ remain hot until $z=0$, while new stars reform a cold disk in the outer regions $R_e<r<5R_e$. The cold disk formed after the merger gradually warms over the past $\sim 3$ Gyr, probably due to gas removal and tidal heating by a cluster. One consequence of the major merger is the absence of old stars on cold orbits, and thus the stellar age distribution of the cold disk can serve as an indicator of the epoch of the last massive merger \citep{Zhu2022b}.

In summary, the dynamically hot stars within 1$R_e$ of these galaxies have diverse origins. In low mass galaxies without a significant merger, the hot stars at $z=0$ are predominantly born on hot orbits and have largely preserved their original kinematics. In contrast, in high-mass galaxies, most stars are initially formed on cold orbits and later heated to hotter orbits. This heating can be attributed to secular evolution or merger events, depending on the galaxy's merger history.
Notably, secular evolution heats stars in the inner region, while major mergers create hot stars across the whole galaxy.

    \begin{figure*}
    \centering
    \includegraphics[width=1.7\columnwidth]{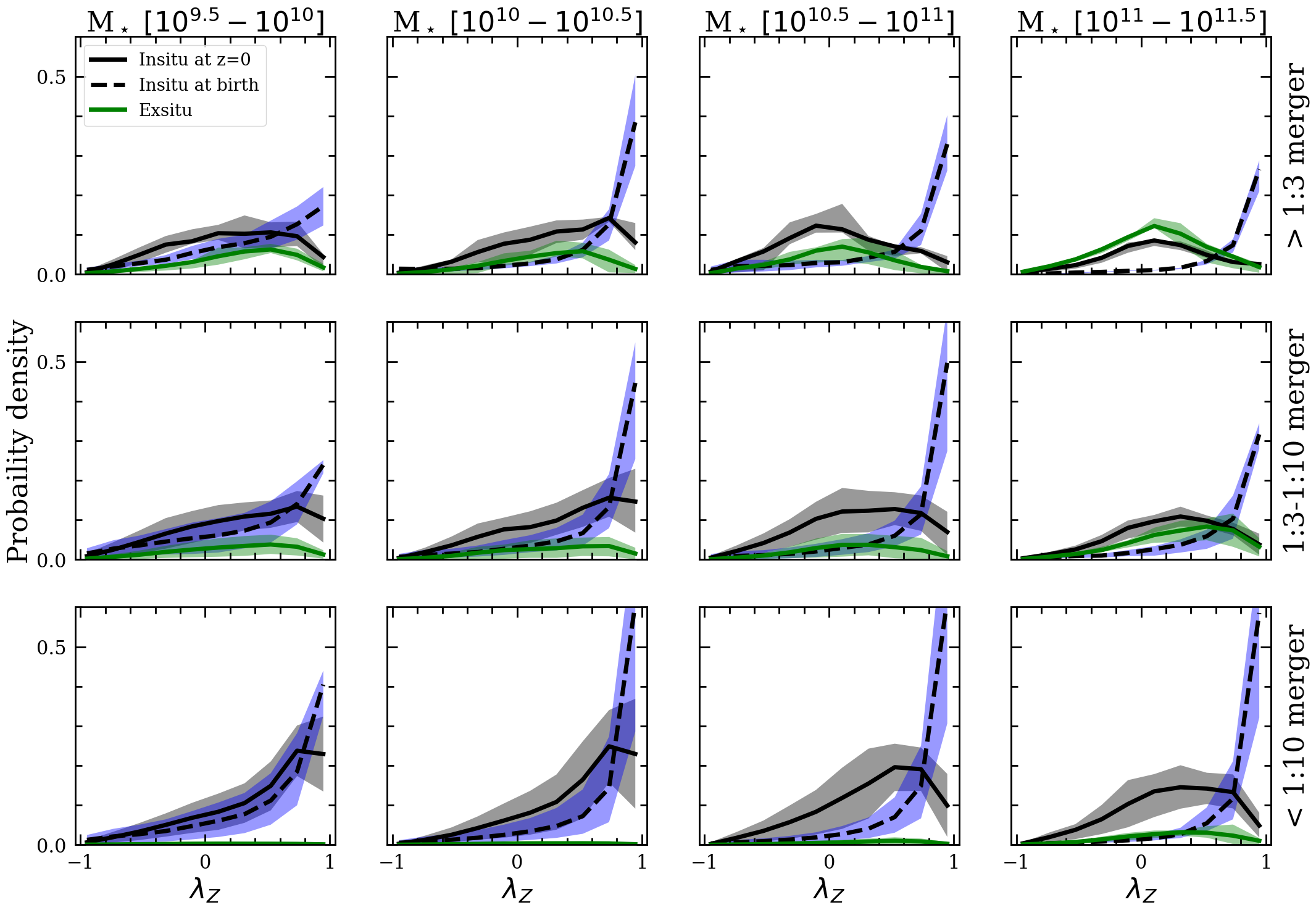}
        \caption{The circularity distribution of stars at $r<R_e$ for galaxies with different stellar mass, akin to the top panel of Fig. ~\ref{fig:bieRe}, but with galaxies categorized into three rows based on their merger histories. From top to bottom, they are galaxies with at least one major merger (mass ratio larger than 1:3), galaxies experienced mergers with ratios between 1:3 and 1:10, and galaxies have not experienced any mergers with a mass ratio larger than 1:10. In lower mass galaxies, the \textit{in-situ} stars are born with a wide distribution of circularity; they are significantly heated in galaxies with major mergers and keep almost unchanged in galaxies with quiescent merger history. In higher mass galaxies, the \textit{in-situ} stars are mostly born cold, especially in galaxies that have not experienced major mergers; all \textit{in-situ} stars at $r<R_e$ are significantly heated, resulting in similar circularity distributions at $z=0$ irrespective of the merger history.}
    \label{fig:bie_merger}
    \end{figure*}

    \begin{figure*}
    \centering
    \includegraphics[width=1.7\columnwidth]{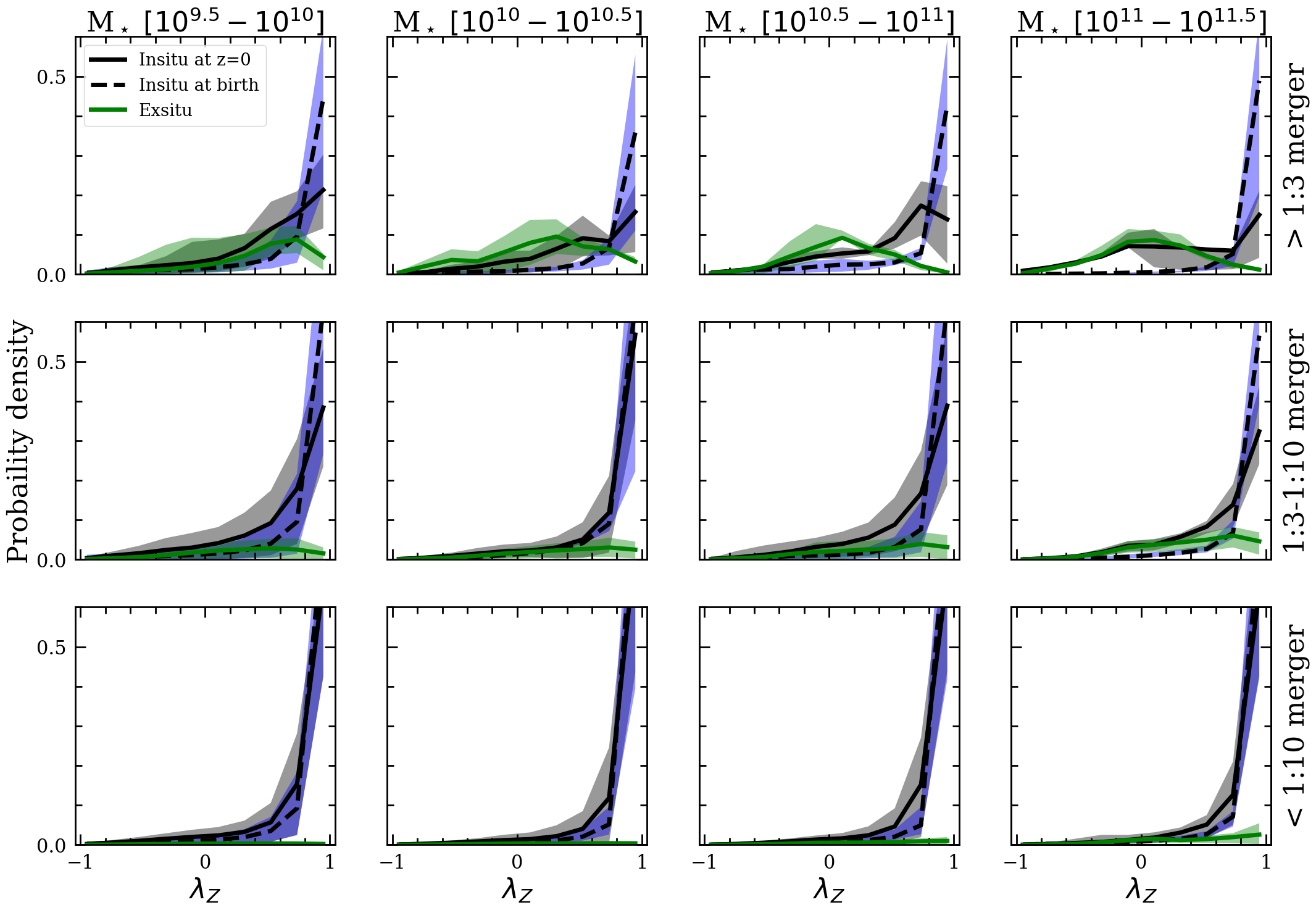}
        \caption{Similar to Fig. \ref{fig:bie_merger} but for stars in the outer regions ($R_e<r<5R_e$). For galaxies in all mass bins, the stars in the outer regions are born cold, especially colder in galaxies with quiescent merger history in which they keep almost unchanged until $z=0$; the stars are significantly heated in galaxies that have experienced more mergers.}
    \label{fig:bie_merger_5Re}
    \end{figure*}

    \begin{figure}
    \centering
    \includegraphics[width=1\columnwidth]{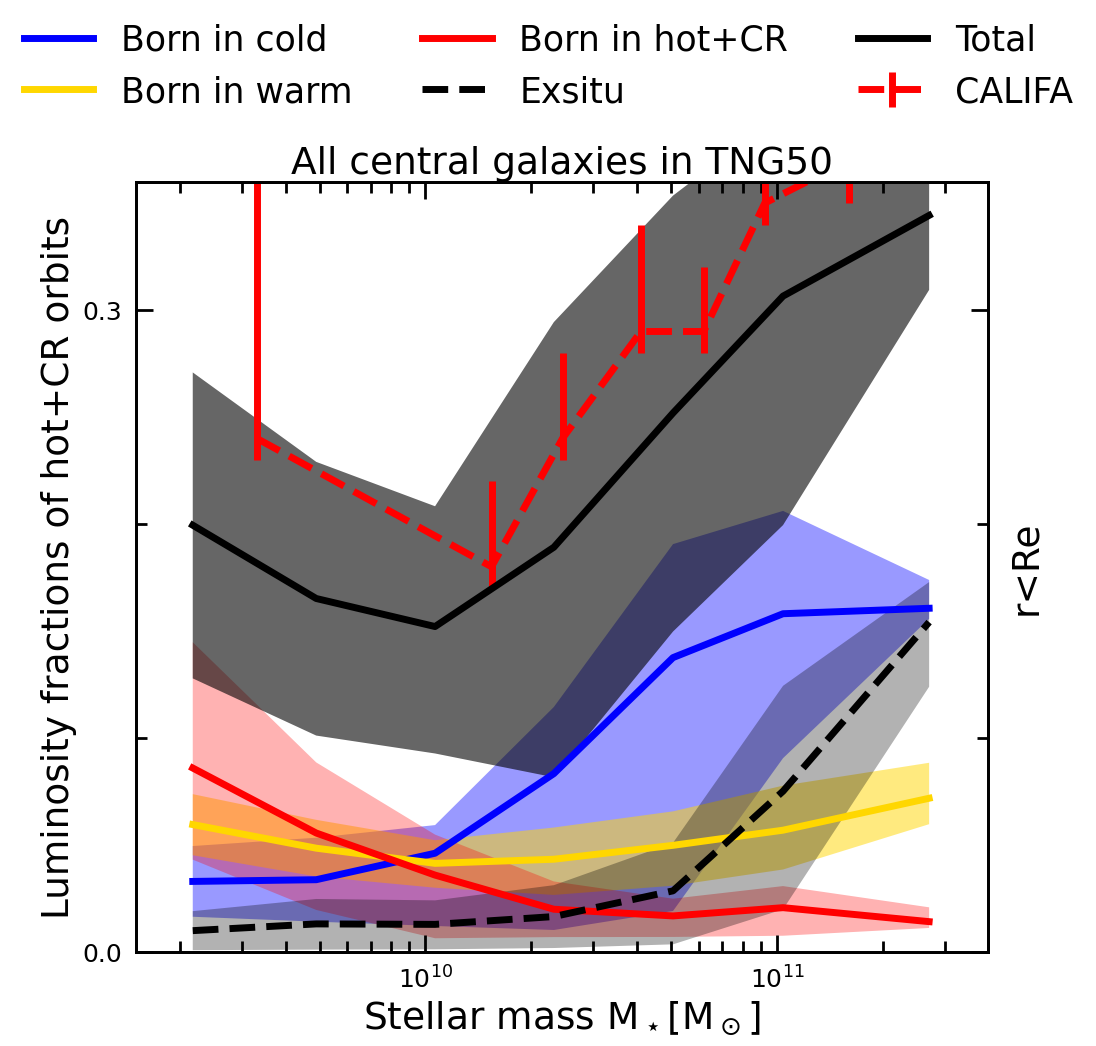}
        \caption{The origin of dynamically hot orbits as a function of stellar mass, encompassing all central galaxies in TNG50. The solid black curve depicts the overall luminosity fraction of hot orbits in galaxies at $z=0$, consistent with the CALIFA galaxies from observations (red dashed). The blue, yellow, and red curves indicate the contribution from \textit{in-situ} stars born cold, warm, and hot, respectively. The dashed curve represents the contribution of \textit{ex-situ} stars. The shadow areas are the $1\sigma$ scatter of galaxies in each mass bin.}
    \label{fig:origin_hot_a}
    \end{figure}

    \begin{figure*}
    \centering
    \includegraphics[width=1.8\columnwidth]{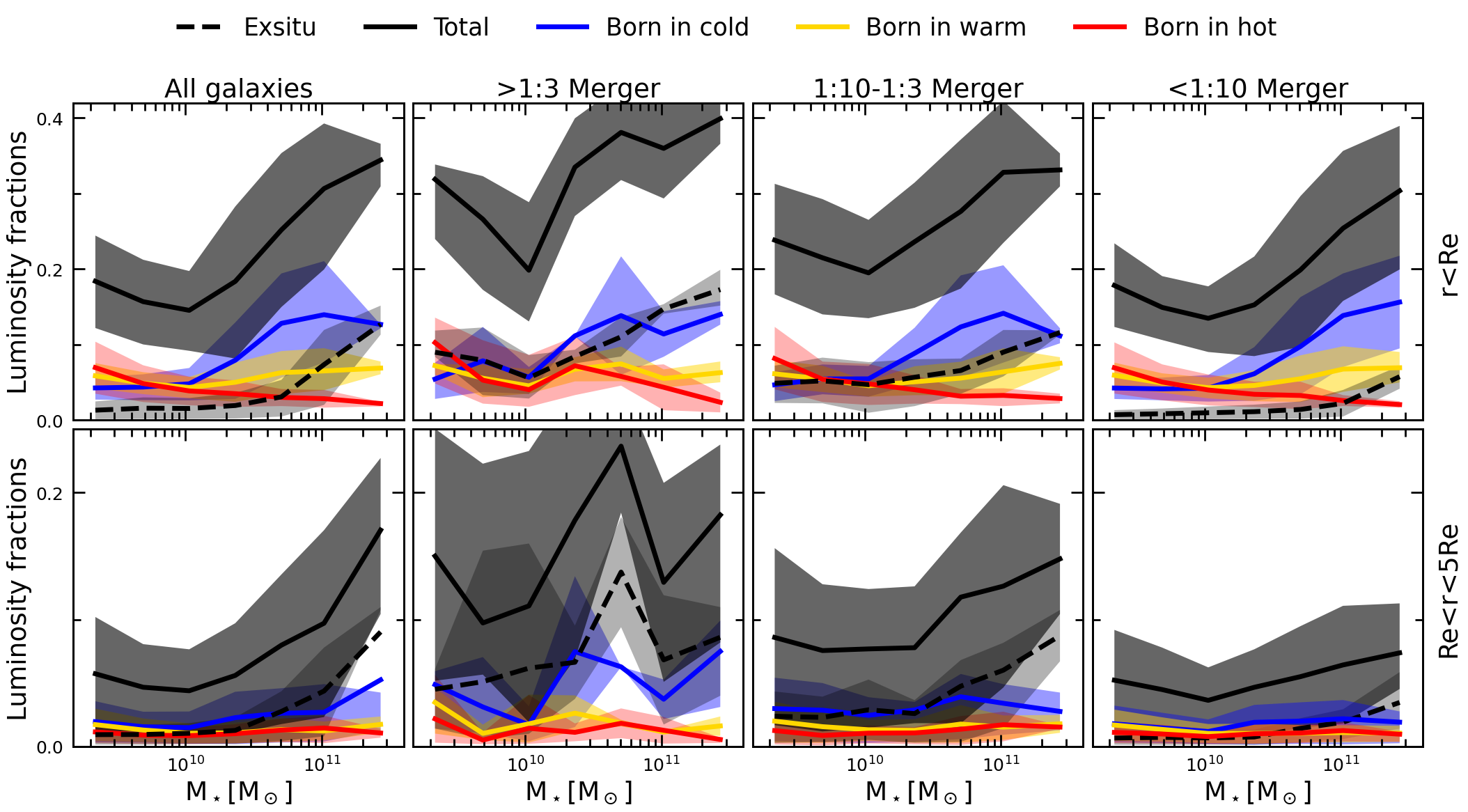}
        \caption{The origin of dynamically hot orbits as a function of stellar mass, for all central galaxies in TNG50, galaxies with major mergers (mass ratio $>1:3$), galaxies with minor mergers (mass ratio $1:3-1:10$), and galaxies without any mergers with mass ratio $>1:10$, from left to right. The stars in the inner regions ($r<R_e$) and the outer regions ($R_e<r<5R_e$) are shown at the top and bottom, respectively. In each panel, the solid black curve depicts the overall luminosity fraction of hot orbits in galaxies at $z=0$, and the blue, yellow, and red curves indicate the contribution from \textit{in-situ} stars born cold, warm, and hot, respectively, the dashed curve represents the contribution of \textit{ex-situ} stars. 
        The dynamically hot stars in the inner regions ($r<R_e$) originate from a combination of different physical processes, and significant hot stars are produced in galaxies even without any mergers. In contrast, hot stars in the outer regions ($R_e<r<5R_e$) are primarily a product of mergers, either accreted or heated through merger processes.}
    \label{fig:origin_hot}
    \end{figure*}

\subsection{The dynamical heating as a function of stellar mass}
\label{subsub:heat}
In the above cases, we have seen that stars may be significantly heated after birth.
We statistically analyse the dynamical heating that occurred in TNG50 central galaxies in a wide range of stellar masses by classifying them into four mass bins with intervals of $M_*=10^{9.5}, 10^{10}, 10^{10.5}, 10^{11}, 10^{11.5}$ \Msun. For galaxies at $z=0$, we separate \textit{in-situ} and \textit{ex-situ} stars, and for \textit{in-situ} stars, we trace their formation history and find their orbital circularity at birth, as described in Section~\ref{sec:Methodology}. 

In Fig. \ref{fig:bieRe}, we show the circularity distribution of \textit{in-situ} and \textit{ex-situ} stars in present-day galaxies at $z=0$, as well as the birth circularity distribution of the \textit{in-situ} stars, for all TNG50 central galaxies. Galaxies in different stellar mass bins are shown from left to right. The stars in the inner ($r<R_e$) and outer regions ($R_e<r<5R_e$) are depicted in the top and bottom panels, respectively, to show the different effects of dynamical heating on the inner and outer regions of galaxies.

For the inner regions ($r<R_e$), stars in lower mass galaxies ($M_*<10^{10}$\ \Msun) are born with a wide distribution of $\lambda_z$, and keep the distribution almost unchanged until $z=0$. On the other hand, stars in higher mass galaxies ($M_*>10^{10}$\ \Msun) are born on dynamically colder orbits with a pronounced peak at $\lambda_z \sim 1$; these stars undergo significant heating and become dynamically hot by $z=0$, with the heating more important in more massive galaxies. In the most massive galaxies, the \textit{in-situ} stars at $z=0$ have a circularity distribution centralised around $\lambda_z \sim 0$, similar to that of \textit{ex-situ} stars. The \textit{ex-situ} stars in the $z=0$ galaxies are mostly dynamically hot but only contribute a significant fraction of stars in the most massive galaxies.

In the outer regions ($R_e<r<5R_e$), \textit{in-situ} stars are predominantly born on dynamically cold orbits across all mass bins. These stars experience only marginal heating until $z=0$. Even in the most massive galaxies, the \textit{in-situ} stars at $z=0$ remain significantly colder than the \textit{ex-situ} stars.

To further understand the role of mergers in dynamical heating, for galaxies in each mass bin, we further separate them into three categories based on their merger histories: galaxies with at least one major merger with mass ratio $>1:3$ (27 galaxies in our sample), at least one minor merger with mass ratio $1:3-1:10$ (246) or only mini mergers with a mass ratio of $<1:10$ (1565).

In Fig. ~\ref{fig:bie_merger}, we show the stars in the inner regions ($r<R_e$) for galaxies in different stellar mass bins from left to right, and with different merger histories from top to bottom. In lower mass galaxies, the circularity distribution of the \textit{in-situ} star remains almost unchanged in galaxies with quiescent merger history, while noticeably heated in galaxies with major mergers; mergers play a significant role in heating stars. In higher-mass galaxies, stars in all galaxies are substantially heated, irrespective of their merger histories. In the most massive galaxies, even galaxies that have not experienced any mergers above 1:10, the \textit{in-situ} stars at $z=0$ are dynamically hot, with circularity distribution centered at $\lambda \sim 0$, similar to the \textit{ex-situ} stars. This suggests that in massive galaxies, stars in the inner $R_e$, if not heated by mergers, will be anyway heated by secular evolution. The heating processes occur universally for all galaxies in the inner regions, and thus the dynamically hot bulge in the inner $R_e$ is not a good indicator of merger history.

In Fig. ~\ref{fig:bie_merger_5Re}, we show a similar figure for stars in the outer regions ($R_e<r<5R_e$). In galaxies in all massive bins, the \textit{in-situ} stars in the outer regions remain dynamically cold until $z=0$ if they did not experience significant mergers or only experienced mini mergers; they are only heated in galaxies that experienced massive mergers. The \textit{ex-situ} stars in the outer regions are also dynamically hot. The dynamically hot stars in the outer regions are mainly produced by mergers, either accreted or heated.
This suggests that to understand a galaxy's merger history, observational data should cover its outer regions.

\subsection{Quantifying the diverse origins of hot bulge}
\label{subsub:origin}
We have learnt that stars of a galaxy form dynamically in distinct ways and are heated by various physical processes, depending on the galaxy's mass and merger history. Here we focus on quantifying the physical origins of the dynamically hot component in galaxies at $z=0$. We trace the formation history of each star in the dynamically hot component and characterise its origin as \textit{in-situ} born cold, born warm, born hot, or \textit{ex-situ}. Again, the birth circularity of \textit{in-situ} stars is found at the snapshot where the star particle first appeared. 

In Fig. \ref{fig:origin_hot_a}, we show the overall luminosity fraction of the hot component within $R_e$ as a function of stellar mass, for the TNG50 and CALIFA galaxies, the same as that shown in Fig. \ref{fig:Flum}. We further show the luminosity fraction of the hot stars originating \textit{in-situ} as born cold, born warm, born hot, and \textit{ex-situ}.
In lower mass galaxies ($M_*\lesssim 10^{10}$\ \Msun), a significant fraction of hot stars were born hot and remain hot until today. In galaxies with $M_*\gtrsim 10^{10}$ \Msun, the fraction of stars born hot decreases, and the fraction of stars born cold/warm but heated increases with galaxy stellar mass. Only in galaxies with $M_*\gtrsim 5\times 10^{10}$ \Msun, the contribution of \textit{ex-situ} stars becomes more and more important. The combination of these three physical origins results in the U-shaped profile of the overall hot orbit fraction as a function of galaxy stellar mass. 

We further separate galaxies into three groups based on their merger histories: those that experienced $>1:3$ major mergers, $1:10-1:3$ minor mergers, and those only with $<1:10$ mini mergers, and check the origin of hot stars in both the inner ($r<R_e$) and outer regions ($R_e<r<5R_e$), as shown in Fig. \ref{fig:origin_hot}.
For stars in the inner regions ($r<R_e$), the hot orbit fraction as a function of galaxy stellar mass exhibits a similar U-shaped pattern across galaxies with different merger histories. The hot orbit fraction is consistently higher in galaxies with more major mergers, primarily due to the significant contribution of hot stars directly accreted from mergers. The stars born cold but heated show a similar profile of increasing with increasing stellar mass; the inner regions of galaxies are anyway heated by secular evolution, if not by mergers. 

In summary, the dynamically hot component in the inner $R_e$ of present-day galaxies has various physical origins, in terms of galaxy stellar mass and merger history. The combination of these physical processes results in the U-shaped profile of the hot orbit fraction in the inner $R_e$ as a function of stellar mass, similar to what is observed. The dynamically hot bulge in the inner $R_e$ is thus not a good indicator of the galaxy merger history.

In contrast, in the outer regions ($R_e<r<5R_e$), galaxies with major mergers have significantly higher fractions of a hot component than those with quiescent merger histories, and the hot stars are either accreted or heated by the mergers. The dynamically hot stars in the outer regions $R_e<r<5R_e$ can be used to indicate the galaxy merger history. This is consistent with previous studies \citep{Zhu2022a} showing that a hot inner stellar halo, defined by dynamically hot stars ($\lambda_z < 0.5$) at $r>3.5$ kpc, is a strong indicator of galaxies' total \textit{ex-situ} stellar mass.

\subsection{The evolution of galaxy structures with redshift}

    \begin{figure*}
    \centering
    \includegraphics[width=1.8\columnwidth]{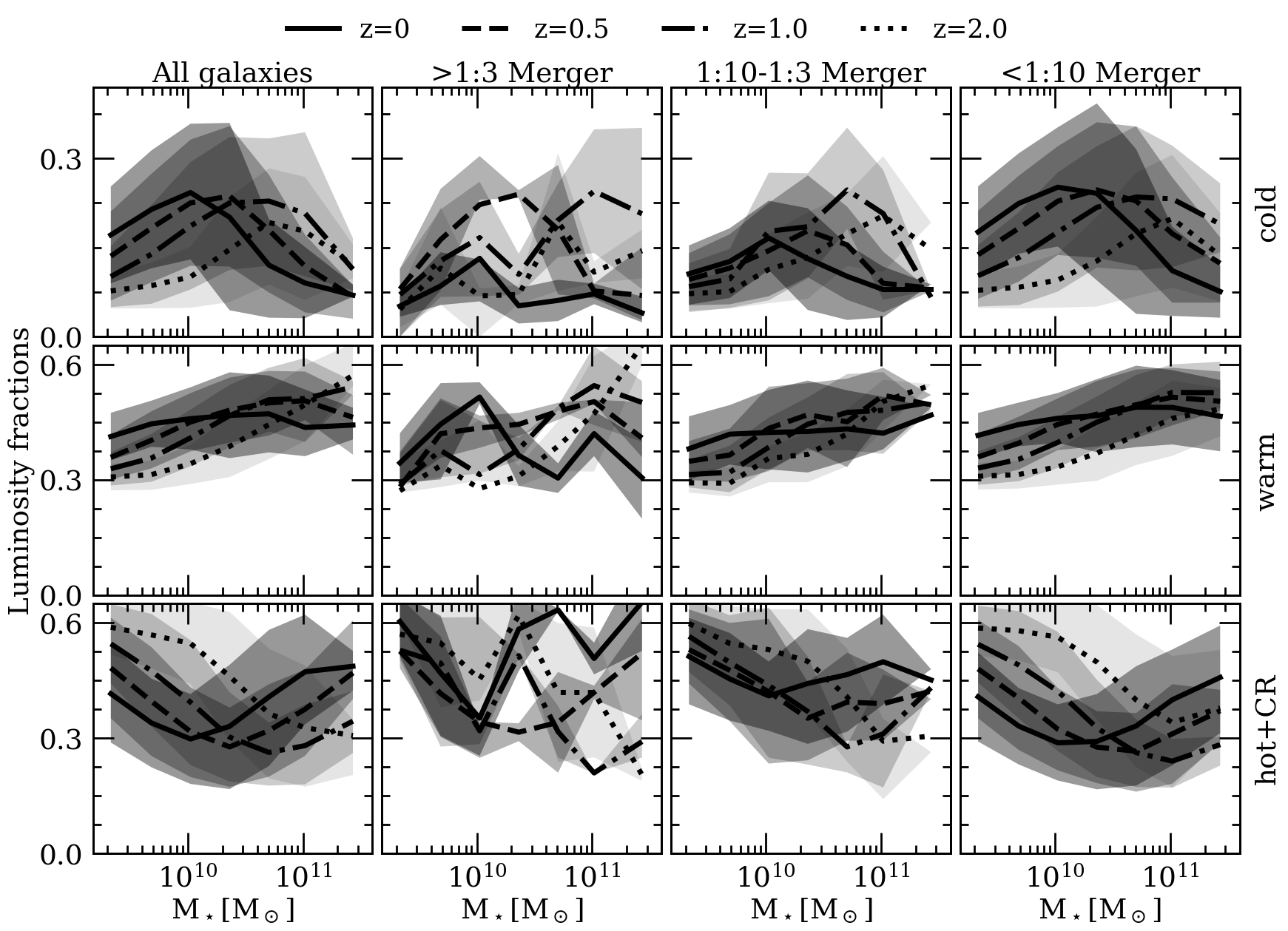}
        \caption{The evolution of galaxy orbital composition with redshift. The columns from left to right are all galaxies, galaxies with major mergers (mass ratio > 1:3), galaxies with minor mergers (mass ratios between 1:10-1:3), and galaxies with negligible mergers or no mergers (mass ratio < 1:10). The rows from top to bottom represent the mass fractions of cold, warm, hot+CR components as functions of the galaxy's stellar mass within $R_e$. In each panel, the solid, dashed, dash-dotted, and dotted curves represent galaxies at $z=0$, and their progenitors at $z=0.5$, $z=1$, and $z=2$, respectively.}
    \label{fig:evo_z}
    \end{figure*}

A galaxy is assembled by all the stars that were formed and by all the dynamical processes that occurred in the past. In this section, we further show how the accumulation of the physical processes we analyse causes the evolution of galaxy structures over time.
We first calculate the mass fraction of the cold, warm, hot, and CR components within $R_e$ as a function of the galaxy stellar mass for all galaxies at $z=0$, similar to Fig. ~\ref{fig:Flum}, but with mass fractions instead of luminosity fractions used here to enable a direct comparison with their progenitors. Then for each galaxy, we find their main progenitor galaxy at $z=0.5$, $z=1.0$, and $z=2.0$, and calculate the mass fraction of the four components also within $R_e$ measured at $z=0$.
 
In Fig. ~\ref{fig:evo_z}, we show the mass fraction of cold, warm, hot+CR components as a function of stellar mass, from top to bottom, together with those in their main progenitors at different redshifts. And from left to right, we show all galaxies, galaxies with major mergers ($>1:3$), galaxies with minor mergers ($1:10-1:3$), and those without any mergers ($<1:10$). 

In general, lower-mass galaxies at $M_*\lesssim 2 \times 10^{10}$ \Msun\, were dynamically hotter in the past, with a lower fraction of cold orbits in their progenitors at higher redshift. This is consistent with the case of the low mass galaxy we show in Section~\ref{subsub:cases}, whose stars are born hotter in the high redshift and colder in the low redshift. The stars remain as they are born in the cosmic timescale. Dynamical heating does not play a major role in these galaxies.
In contrast, the higher-mass galaxies $M_*\gtrsim 2\times 10^{10}$ \Msun\, were dynamically colder in the past, with a higher fraction of cold orbits in their progenitors at higher redshifts. In these massive galaxies, stars were born mostly in cold orbits in all cosmic epochs, and they became more and more heated with time, either by secular heating or mergers.

The structure evolution is different in galaxies with different merger histories, as expected. 
The number of galaxies with major mergers (mass ratio >1:3) that happened in a certain cosmic epoch is small, so the curves in the second column of Fig. \ref{fig:evo_z} are stochastic. However, in general, galaxies in this group across all mass ranges have once had higher cold-disk fractions in the past than today; they are dynamically hotter at $z=0$ caused by major mergers accumulating across the lifetime. 
For galaxies with quiescent merger histories (mass ratio <1:10), their structure evolution is not significantly altered by mergers. Thus, the low-mass galaxies smoothly change from hot to cold, and the high-mass galaxies change from cold to hot from high to low redshift. Galaxies with minor mergers (mass ratio 1:3-1:10) experienced structure evolution between the above two groups. 

The structure assembly of the entire galaxy sample statistically aligns with the group of galaxies without any mergers, albeit with a slightly different peak galaxy mass exhibiting the maximum cold orbit fraction at $z=0$. This is expected in low-mass galaxies with low merger frequencies. The merger frequency is higher in higher mass galaxies, and mergers create significant hot orbits in the most massive galaxies; however, when accumulated with lifetime, secular heating can anyway heat the disk within $R_e$ in the absence of mergers, leading to a statistically similar evolution of orbital structures in galaxies without mergers at the high mass end. Note that we are only considering stars within $R_e$ here, if we consider all stars including those in the outer regions, the structure evolution should be different for the massive galaxies with or without mergers.

The structure evolution of massive galaxies shown here is consistent with that found from the EAGLE simulations \citep{Santucci2024}, and also with the direct observations at high redshift. 
The compact star-forming galaxies presenting SFR = 100-200 \Msun $yr^{-1}$ observed at high redshifts are mostly rotational-supported, and they are supposed to be the progenitors of quiescent galaxies in the local universe \citep{wisnioski2018kmos3d, van2015forming, barro2013candels, barro2015extreme}.
ETGs at $z\sim 1$ were also found to be more rotation-supported than their local counterparts \citep{Bezanson2018, DEugenio2023, stott2016kmos, forster2018witnessing}.

\section{Discussion}
\label{sec:discuss}

\subsection{Limitations in observations and simulations}
The comparison between the stellar orbit distributions in TNG100, TNG50, and CALIFA galaxies provides valuable insights, and it is crucial to consider certain factors when interpreting the differences.

Given the resolution and scale of our samples, the most statistically robust comparison emerges within the intermediate mass range $M_*$ $\in$ [$10^{10}$,$10^{11}$] \Msun, where the three datasets exhibit strong consistency in the luminosity fractions of the four orbital components. The TNG model adeptly replicates the dynamics of the real universe and remains largely unaffected by resolution disparities.

In low mass galaxies with $M_*\lesssim 10^{10}$ \, \Msun, TNG50 galaxies exhibit a higher cold-disk fraction compared to CALIFA. However, it is important to note limitations in CALIFA's kinematic maps for galaxies in this mass range. Some CALIFA galaxies have limited spatial bins in their kinematic maps, and the spectral resolution of CALIFA may not be sufficient to fully resolve the velocity dispersion of these galaxies \citep{falcon2017}, although the peculiar high-dispersion data point is carefully excluded \citep{zhu2018nature}. This limitation could introduce biases in the kinematic maps, potentially impacting the comparison with TNG50. Therefore, before definitive conclusions are drawn, further confirmation from IFU data with robust kinematics is necessary for low-mass galaxies.

For the most massive galaxies with $M_* \gtrsim 2\times 10^{11}$ \,\Msun, TNG simulations suggest more warm and fewer hot orbits compared to CALIFA. This discrepancy is attributed to the orbit classification method used in the simulations. For the simulated galaxies, we employ phase-space averaging rather than precise orbit integration to construct the circularity of orbits for stellar particles (\ref{subsec:decomposition}). Although this method is effective in narrowing down the circularity distribution of particles in highly radial orbits, it is not perfect. Some particles in highly radial orbits may still exhibit relatively large $\lambda_z$ after phase space averaging, leading to being classified as warm or CR components. The most massive galaxies, dominated by highly radial orbits, are particularly affected by this issue.

The morphologies of the cold, warm, and hot + CR components in TNG50 and TNG100 galaxies are generally consistent with those of CALIFA galaxies, except for a notable difference in the concentration of warm and hot components. This deviation is associated with a known limitation in TNG50, where the simulation struggles to accurately capture the dense inner regions of quenched spheroidal galaxies \citep{Zanisi2021}.

We find a good consistency between the CALIFA and TNG galaxies of a U-shaped profile of the hot orbit fraction as a function of stellar mass, with a minimum at $M_* \sim 1-2\times 10^{10}$\,\Msun. This indicates that both the observations and the simulations support the diverse origins of the dynamically hot bulges in different galaxies across a wide mass range. Such U-shaped profiles are also found in EAGLE \citep{Santucci2024}. For galaxies with or without mergers, we presume a similar diverse origin of hot orbits in EAGLE.

However, it is not guaranteed that the physical origins we find in the TNG simulations all truly occur in real galaxies. The formation of dynamically hot bulges is still highly degenerate in simulations with complicated feedback models and dynamical processes. Taking MW-like galaxies (galaxies with $M_{*} \sim 5\times10^{10} M_{\odot}$) as an example, the formation of dynamically hot bulges are different in different simulations. In TNG50, we find that stars are mostly born cold in MW-like galaxies, and they are heated mostly by secular evolution to form a dynamical hot bulge (like the second case in Fig. ~\ref{fig:merger}). In FIRE-2 with strong stellar feedback, stars formed in a bursty phase at high redshift, they remain unchanged and form the dynamically hot bulge at $z=0$ \citep{Yu2021}. In Auriga, secular heating by the bar is also an important process in heating the stars and creating an \textit{in-situ} bulge for MW-like galaxies if no significant mergers. It is hard for us to constrain the formation of bulges further with the luminosity fractions measured from observations; the chemodynamical properties of the stars in the bulge might help.

\subsection{Dynamical heating in secular evolution}

We find that stars in the inner regions ($r<R_e$) of massive galaxies with $M_{*}\gtrsim 10^{10}$\ \Msun are universally heated in all galaxies, irrespective of their merger histories, with small scatters, as shown in Fig. \ref{fig:bie_merger}. 

We consider secular evolution to play a dominant role in producing dynamically hot bulges in massive galaxies that do not experience massive mergers with a mass ratio $>1:10$.
The merger mass ratio is defined by the stellar mass of the two progenitor galaxies. We might miss some mergers with dark galaxies (with very few or no stars) or galaxy interactions, which can also heat the stellar disk of the main progenitor galaxy. However, heating by mergers should have a stronger or at least similar impact on the outer disk (see \ref{subsub:cases}).
However, the outer disks of these galaxies remain cold, as shown in Fig. \ref{fig:bie_merger_5Re}. Such mergers with dark galaxies should not be a major reason for the heating of the inner regions either. For similar reasons, interactions with other galaxies, such as fly-by, can not be a major reason either.

    \begin{figure}
    \centering
    \includegraphics[width=1\columnwidth]{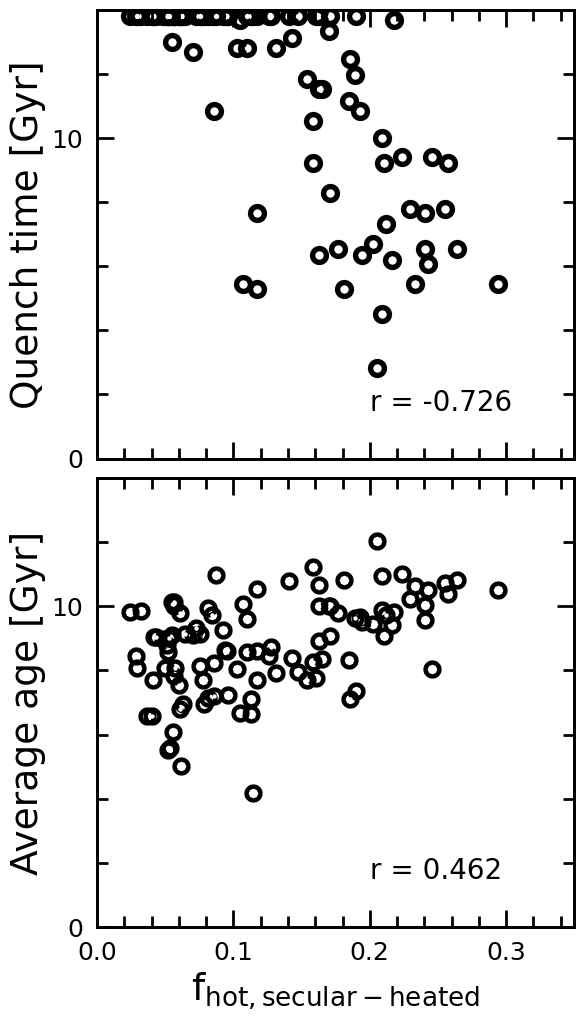}
        \caption{The luminosity fraction of hot orbits heated by secular evolution $f_{\rm hot,secular-heated}$ vs. the cosmic time when the galaxy quenched (upper panel), and vs. the average stellar age of this group of stars (lower panel). All TNG50 massive galaxies (stellar mass $M_*\ge 10^{10.5}$ \Msun) with quiescent histories (merger ratio of < 1:10) are included. For galaxies that are still with star formation at $z=0$, we set quench time to be the age of the universe (13.8 Gyr). The Pearson correlation coefficients of the two panels are labeled. Galaxies quenched at an earlier stage are more significantly heated by secular processes, while the average ages of their stars exhibit a weaker correlation.}
    \label{fig:qt_y_age}
    \end{figure}

Such secular evolution is found for the formation of massive quiescent galaxies in TNG50 \citep{park2022formation}; half of such galaxies at $z=0$ are rapidly quenched at higher redshift while still disk-like, and they become more elliptical mostly by disk heating but still maintain some degree of rotation at $z=0$. This is consistent with what we find that only the inner regions of galaxies are significantly heated by secular evolution, not the outer regions. \citet{park2022formation} also found that quenching and morphological transformation are largely decoupled in these galaxies; the morphological transformation appears to be not directly correlated with the feedback processes. 

In our analysis, secular evolution has occurred in almost all massive galaxies without major mergers, irrespective of their star formation history. 
In Fig. ~\ref{fig:qt_y_age}, we show all massive galaxies ($M_*\ge 10^{10.5}$) that have only experienced mini mergers with the merger mass ratio $<1:10$. We take the luminosity fraction of hot orbits born cold to represent stars heated by secular evolution $f_{\rm hot,secular-heated}$ because mergers do not play a significant role in heating the cold orbits in these galaxies.
As shown in Fig. ~\ref{fig:qt_y_age}, the luminosity fraction of stars heated by secular evolution $f_{\rm hot,secular-heated}$ is correlated with the quench time of the galaxy, with the Pearson's correlation coefficient $r = -0.726$, stronger than the correlation between $f_{\rm hot,secular-heated}$ and the age of the stars that have been heated. Although some star-forming galaxies have a high fraction of stars heated through secular evolution, the disks of early quenched galaxies are more inclined to be heated. One possible explanation is that the rapid depletion of gas by supermassive black holes (SMBHs) \citep{varma2022building} in the central region reduces the stability of the stellar disk, making it more susceptible to bar heating \citep{grand2016vertical} or interactions with flyby galaxies. Furthermore, \citet{walters2021structural} suggests that the accretion history of SMBHs can indicate rapid core growth: galaxies with higher SMBH accretion rates exhibit faster increases in core density. The processes of quenching and disk heating might be decoupled, although the timing of quenching can influence subsequent transformations.

The heating by secular evolution might be correlated with the onset of bars in the simulation, caused by either internal instability or external interactions.
There is a high fraction of barred galaxies in massive galaxies of TNG and EAGLE, which could be as high as $f_{\rm bar} \sim 70\%$ in galaxies with $M_*\sim 10^{11}$ \Msun\ \citep{Roshan2021}, although it could be much lower (with $f_{\rm bar} \sim 30\%$) with different selection criteria \citep{Zhao2020ApJ...904..170Z}. Some barred galaxies undergo episodes of bar creation, destruction, and regeneration in cosmological simulations \citep{Cavanagh2022MNRAS.510.5164C}, thus the bar heating might be even more common than the fraction of barred galaxies found at $z=0$. Bars expel angular momentum to dark matter in resonances \citep{athanassoula2005bars}, as well as contribute to the development of boxy/peanut-shaped bulges \citep{athanassoula2016boxy, anderson2024interplay}.
The latest study in \citet{lu2024illustristng} shows that TNG produces bars reasonably well, though there may be somewhat more numerous short bars in TNG50 than observed \citep[see also][]{Frankel2022}. It is noticed that the growth of bars drives the decrease of the overall rotation of disk galaxies, which partially contributes to the dynamical heating in the secular evolution in our study.

    \begin{figure}
    \centering
    \includegraphics[width=1\columnwidth]{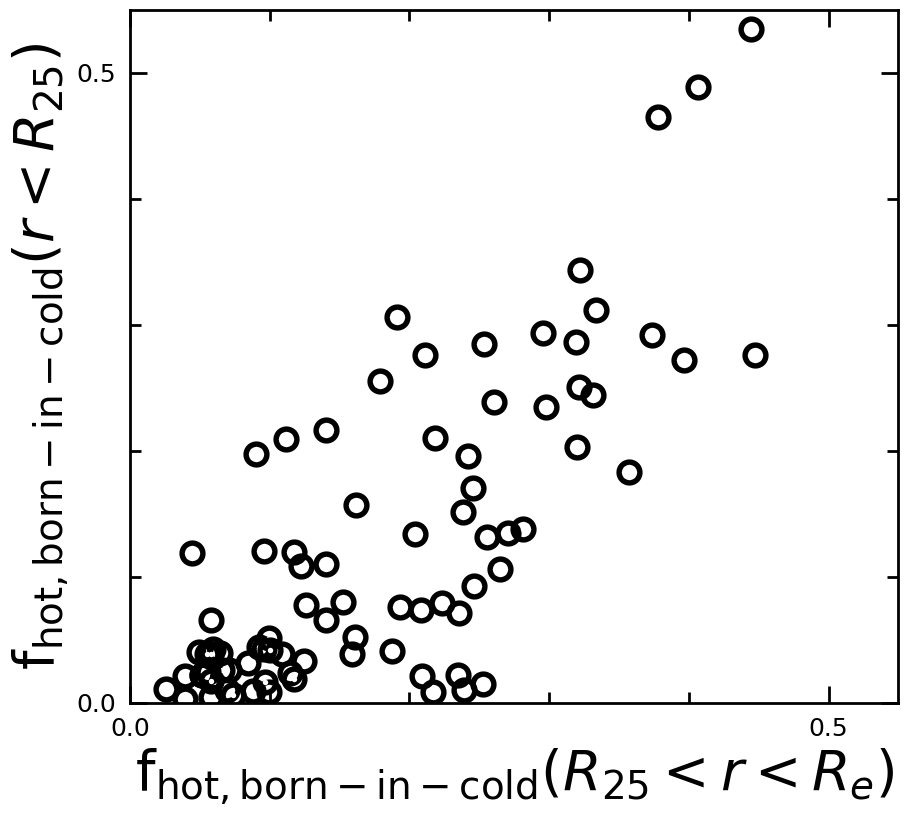}
        \caption{The luminosity fractions of hot stars with cold origins at different radius, $R_{25}$<r<$R_e$ (x-axis) vs. r<$R_{25}$ (y-axis), for the massive galaxies (stellar mass $M_* \ge 10^{10.5}$ \Msun) with quiescent histories (merger ratio of $< 1:10$), the same galaxy sample as Fig. \ref{fig:qt_y_age}. Secular heating exerts a similar effect on orbits for $R_{25}$<r<$R_e$ and r<$R_{25}$, indicating that spurious heating does not play a major role.}
    \label{fig:spurious}
    \end{figure}  

Due to the limited resolution of the simulations and different mass resolutions of stellar and dark matter particles, stars in simulation could be spuriously heated by numerical reasons \citet{ludlow2023spurious}. In our analysis, hot stars in the low mass galaxies are mostly born hot, and spurious heating should not be a major factor of their origins. But stars in massive galaxies are mostly born cold and heated later. To check the importance of spurious heating, we still focus on massive galaxies with $M_{*}\gtrsim 3\times 10^{10}$\ \Msun\ with extremely clear merger histories that have only experienced mergers with the ratio $<1:10$; mergers do not play a significant role in heating, but their inner regions are still gradually heated.

According to the analysis in \citet{ludlow2023spurious}, spurious heating mainly takes place in the galaxy center and mainly influences galaxies in low-mass halos. Above a halo mass of a few $\times 10^{11}$\ \Msun\, the global galaxy structure and kinematic quantities within $R_e$ are largely unbiased by spurious heating. Since spurious heating becomes insignificant only when the number of enclosed particles reaches a minimum threshold, most TNG50 galaxies with $M_{*}\gtrsim 10^{10.5}$\ \Msun\, (halo mass $> 10^{12}$\ \Msun\,) should be safe at the radius enclosing 25 percent of the stellar mass ($R_{25}$).
We compare the prevalence of heating within $r<R_{25}$ and $R_{25}<r<R_e$ in galaxies with $M_{*}\gtrsim 10^{10.5}$\ \Msun\ that have experienced only minor or no mergers (mass ratio $<1:10$). As shown in Fig. \ref{fig:spurious}, the fraction of hot stars formed in cold regions is similar for $r<R_{25}$ and $R_{25}<r<R_e$. Although quantifying the contribution of spurious heating to the heating of stars in these regions remains challenging, it is unlikely that it will be the dominant process.

\section{Conclusions}
\label{sec:conclusion}
In this study, we select a comparative sample of galaxies from the IllustrisTNG simulations (TNG50 and TNG100) that is akin to the CALIFA sample, which in turn comprises 260 field galaxies in the local universe, with a stellar mass range of $M_*$ $\in$ [$10^{9}$,$10^{12}$] \Msun. As shown in \ref{subsec:TNG50}, the simulated samples encompass central galaxies at $z = 0$, with stellar mass ranging $M_* \in [10^{9.7},\,10^{12}]$ \Msun\ for TNG100 and $M_*$ $\in$ [$10^{9}$,$10^{12}$] \Msun\ for TNG50. We begin by decomposing the simulated galaxies into four stellar orbital components -- cold, warm, hot, and counter-rotating components -- based on the stellar circularity distribution of each galaxy, in a way comparable, but not identical, to what is done for CALIFA galaxies with the orbit-superposition method in \citet{zhu2018nature}. We can hence directly compare the luminosity fractions and morphological properties of each orbital component in CALIFA and TNG galaxies to assess the realism of the simulated galaxies and to gain insights from the simulations about the origin of such components.

Our analysis reveals a broad consistency between the galaxies observed in CALIFA and those simulated within TNG (see Fig. \ref{fig:Flum} - Fig. \ref{fig:sersic}). Such an agreement supports the usage of the TNG50 output to trace the stars in the present-day galaxies back to their circularity at birth and to quantify the physical origins of stars in different components; hence, statistically. In this paper, we focus especially on the dynamically hot bulge in the present-day galaxies. 

The direct comparison between CALIFA and the TNG50/TNG100 galaxies reveals several key findings:

\begin{enumerate}
\item Both TNG100 and TNG50 reproduce the typical, i.e. average luminosity fractions of the cold, warm, hot, and counter-rotating components seen in CALIFA galaxies across the stellar mass range of $M_* = 10^9-10^{12}$\ \Msun (Fig. \ref{fig:Flum}). They have a maximum (minimum) cold (hot) orbit fraction at the stellar mass $M_*=1-2\times 10^{10}$\ \Msun, whereby the cold orbit fraction decreases and the hot orbit fraction increases in galaxies with lower and higher mass. Despite the different underlying resolutions, the TNG100 and TNG50 simulations demonstrate notable consistency with each other.

\item TNG100 and TNG50 galaxies broadly reproduce the morphological properties of each component found in the CALIFA galaxies (Fig. \ref{fig:qe}). The intrinsic flattening within $R_e$ remains nearly constant for each component, with values $\sim 0.3, 0.5, 0.9$ for the cold, warm, and hot components, respectively, in both CALIFA and TNG galaxies. However, it should be noted that the hot components in TNG galaxies do not exhibit the systematic flattening seen in lower mass ($M_*\lesssim 10^{10}$\ \Msun) CALIFA galaxies.

\item The S\'ersic index of the cold components can be less than 1 for all galaxies, while those of the warm and hot components increase from low to high mass galaxies (Fig. \ref{fig:sersic}). However, the S\'ersic indices of warm and hot components in TNG galaxies are systematically lower than those in CALIFA. i.e., the dynamically warm and hot components in TNG galaxies are not compact enough compared to observations.
\end{enumerate}

By tracing back the formation histories of galaxies in TNG50, we find the following.

\begin{enumerate}
\item Stars on dynamically hot orbits within the inner $R_e$ exhibit varied origins depending on the stellar masses and merger histories of galaxies. In contrast, stars in outer regions ($r>R_e$) are predominantly born on cold orbits and are subsequently heated by mergers. One direct consequence is that the hot orbit fraction in the inner $R_e$ of current galaxies does not strongly correlate with their merger histories, while the hot orbit fraction in the outer regions serves as a reliable indicator of merger history, consistent with previous studies \citep{Zhu2022a, du2021evolutionary}.

\item There are three main regimes and physical origins contributing to the hot orbits in the inner $R_e$ of galaxies (Fig. \ref{fig:merger}):
(1) in the lower mass galaxies ($10^9 < M_* \lesssim 10^{10}$\ \Msun) without significant mergers, hot stars are primarily born on hot orbits: stars in the inner $R_e$ are born with a wide distribution of $\lambda_z$, and generally preserve their $\lambda_z$ distribution until $z=0$; (2) in the higher mass galaxies ($M_*\gtrsim 10^{10}$\ \Msun) without significant mergers, stars are mainly born on cold orbits but undergo prolonged secular heating over cosmic time; (3) mergers significantly increase the hot orbit fraction across all studied mass ranges through the heating of previously cold disks, the accretion of stars from the merging galaxy, and triggering star formation in dynamically hot orbits during the mergers if gas-rich. 

\item The combination of the three physical origins effectively explains the peak in the cold orbit fraction and the dip in the hot orbit fraction in galaxies with $M_*\sim 1-2\times 10^{10}\,$\Msun. In lower mass galaxies, the increase in hot orbit fraction is due to a higher proportion of stars born on hot orbits. In higher mass galaxies, the rise in the hot orbit fraction results from increased heating through secular evolution or mergers, along with a higher fraction of \textit{ex-situ} stars that are mostly in dynamically hot orbits.

\item Secular heating plays a statistically important role in creating dynamically hot stars in the inner $R_e$ for massive galaxies ($M_* \gtrsim 10^{10}\,$\Msun). For these galaxies with quiescent merger histories, all stars are born on cold orbits but are significantly heated until $z=0$, although the circularity $\lambda_z$ distribution is not centered exactly at $\lambda_z=0$, slightly different from the resulting $\lambda_z$ distribution from major mergers. Massive galaxies are mostly born cold and are commonly heated by a long-term secular process, consistent with the early onset of bar, as observed by JWST in galaxies at $z>1$ \citep{Guo2023, Tsukui2023,conte2023jwst}.
\end{enumerate}

In conclusion, according to the TNG simulations, the assembly histories of orbital structures differ markedly in low- and high-mass galaxies. Lower mass galaxies are predominantly hot in the past and have become more disk-dominated at lower redshifts. In contrast, massive galaxies have been statistically more disk-dominated and rotation-supported in the past, aligning with observational results \citep{Kartaltepe2023, Bezanson2018, DEugenio2023}.

\begin{acknowledgements}
    This work is supported by the Max-Planck Partner's group led by LZ and AP.
    LZ acknowledges the support of the CAS Project for Young Scientists in Basic Research under grant No. YSBR-062 and the National Key R\&D Program of China No. 2022YFF0503403. M.D. acknowledges the support by the Fundamental Research Funds for the Central Universities (No. 20720230015), the Natural Science Foundation of Xiamen, China (No. 3502Z202372006), and the Science Fund for Creative Research Groups of the National Natural Science Foundation (NSFC) of China (No. 12221003). J.F-B acknowledges the support from PID2022-140869NB-100 granted from the Spanish Ministry of Science and Innovation.
\end{acknowledgements}

%
%
\bibliographystyle{aa}
\bibliography{Ref}

\end{document}